\documentclass[a4 paper, 11pt]{article}
\pdfoutput=1
\usepackage{jheppub}
\usepackage{subcaption}
\usepackage{amsmath}
\usepackage{amssymb}
\usepackage{amscd}
\usepackage{dsfont}
\usepackage{enumerate}
\usepackage{amsfonts}
\usepackage{epsfig}
\usepackage{mathtools}
\usepackage{yfonts}
\usepackage{bbold}
\usepackage{breqn}
\usepackage[utf8]{inputenc}
\usepackage[english]{babel}

\def\bal#1\eal{\begin{align}#1\end{align}}
\def\alp[#1]{\begin{align}#1\end{align}}
\def\secnum[#1]{\texorpdfstring{$#1$}{TEXT}}
\def\secnuml#1\secnumr{\texorpdfstring{$#1$}{TEXT}}

\def\eqa{\begin{eqnarray}}
\def\eqae{\end{eqnarray}}
\def\eq{\begin{equation}}
\def\eqe{\end{equation}}
\def\be{\begin{equation}}
\def\ee{\end{equation}}
\def\bea{\begin{eqnarray}}
\def\eea{\end{eqnarray}}
\def\ba{\begin{array}}
\def\ea{\end{array}}
\def\bd{\begin{displaymath}}
\def\ed{\end{displaymath}}

\def\ie{{\it i.e.~}}

\def\>{\rangle}
\def\<{\langle}

\def\t{\tau}

\title{Causal shadow and non-local modular flow: from degeneracy to perturbative genesis by correlation}
\begin{document}
\author{Liangyu Chen}
\author{and Huajia Wang}
\affiliation{Kavli Institute for Theoretical Sciences, University of Chinese Academy of Sciences, \\ Beijing 100190, China}
\emailAdd{chenliangyu19@mails.ucas.ac.cn}
\emailAdd{wanghuajia@ucas.ac.cn} 

\abstract{Causal shadows are bulk space-time regions between the entanglement wedges and the causal wedges, their existence encodes deep aspects of the entanglement wedge reconstruction in the context of subregion duality in AdS/CFT. In this paper, we study the perturbation theory of the causal shadows and their relation to the properties of the associated modular flows. We first revisit the cases of degenerate causal shadows based on known examples, and discuss the origin for their degeneracy via the local nature of the modular flow. We then focus on the perturbative case in which the CFT subregion consists of two spheres separated by a large distance $L\gg R_{1,2}$. The RT surfaces still agree with the causal horizons, giving a degenerate causal shadow classically. We compute the corrections to the quantum extremal surfaces (Q.E.S) from the bulk mutual information, which then give rise to a non-degenerate causal shadow at order $G_N$. We end by discussing the causal shadow perturbation theory more generally, in particular we explore the possibility of extracting the positivity conditions characterizing perturbative causal shadows in the boundary CFTs.}

\maketitle
 
\setcounter{page}{1}
\setcounter{tocdepth}{1}

\tableofcontents
\section{Introduction}
AdS/CFT has become a major framework for understanding quantum gravity in special sectors of asymptotic boundary conditions\cite{Maldacena1999, Witten1998AntideSS,GUBSER1998105}. It is an explicit manifestation of the holographic principle \cite{Susskind:1994vu, tHooft:1993dmi}. A key feature conveyed through AdS/CFT is the notion of an emergent bulk theory of quantum gravity based on ingredients in the boundary conformal field theory (CFT). In principle, this implies that elements and phenomena in the bulk gravitational theory can be ``reconstructed" from the boundary CFTs -- an endeavor generally known as the ``bulk reconstruction". 

In the classical limit $G_N\ll 1$ of the bulk theory, such reconstruction can be decomposed into different aspects according to the perturbative expansion in $G_N$. At the leading order, the bulk geometry dual to a state $\psi$ is directly related to the boundary entanglement entropies $S^\psi_A$ in that state via the Ryu-Takayanagi (RT) formula \cite{Ryu2006}:
\be\label{eq:RT}
S^\psi_A = \frac{\text{Area}(\Sigma_A)}{4G_N} 
\ee 
where $\Sigma_A$ is the bulk minimal surface homologous to a subregion $A$, and thus encodes information of bulk metric.  The covariant extension and quantum corrected version of (\ref{eq:RT}) were subsequently proposed \cite{Hubeny2007,Faulkner:2013ana, Engelhardt2015}. At sub-leading orders, the bulk theory consists of fluctuating fields of the bulk effective quantum field theory (QFT) defined on the classical background geometry.  Bulk reconstruction at this order is realized by constructing local operators in the bulk effective QFTs, e.g. writing them in terms of operators in the boundary CFTs. Given the bulk geometry of the dual state, such reconstruction can be accomplished order by order in perturbation theory in $G_N$ \cite{Hamilton2006}: 
\be
\mathcal{O}_{bulk}(z,x) = \int dy\; K(z,x;y)\mathcal{O}_{bdry}(y) + ... 
\ee 
where the kernel $K$ is related to the bulk-boundary propagators in the background dual to $\psi$. This is generally known as the HKLL method.

\begin{figure}[h]
     \centering
     \includegraphics[scale=0.50]{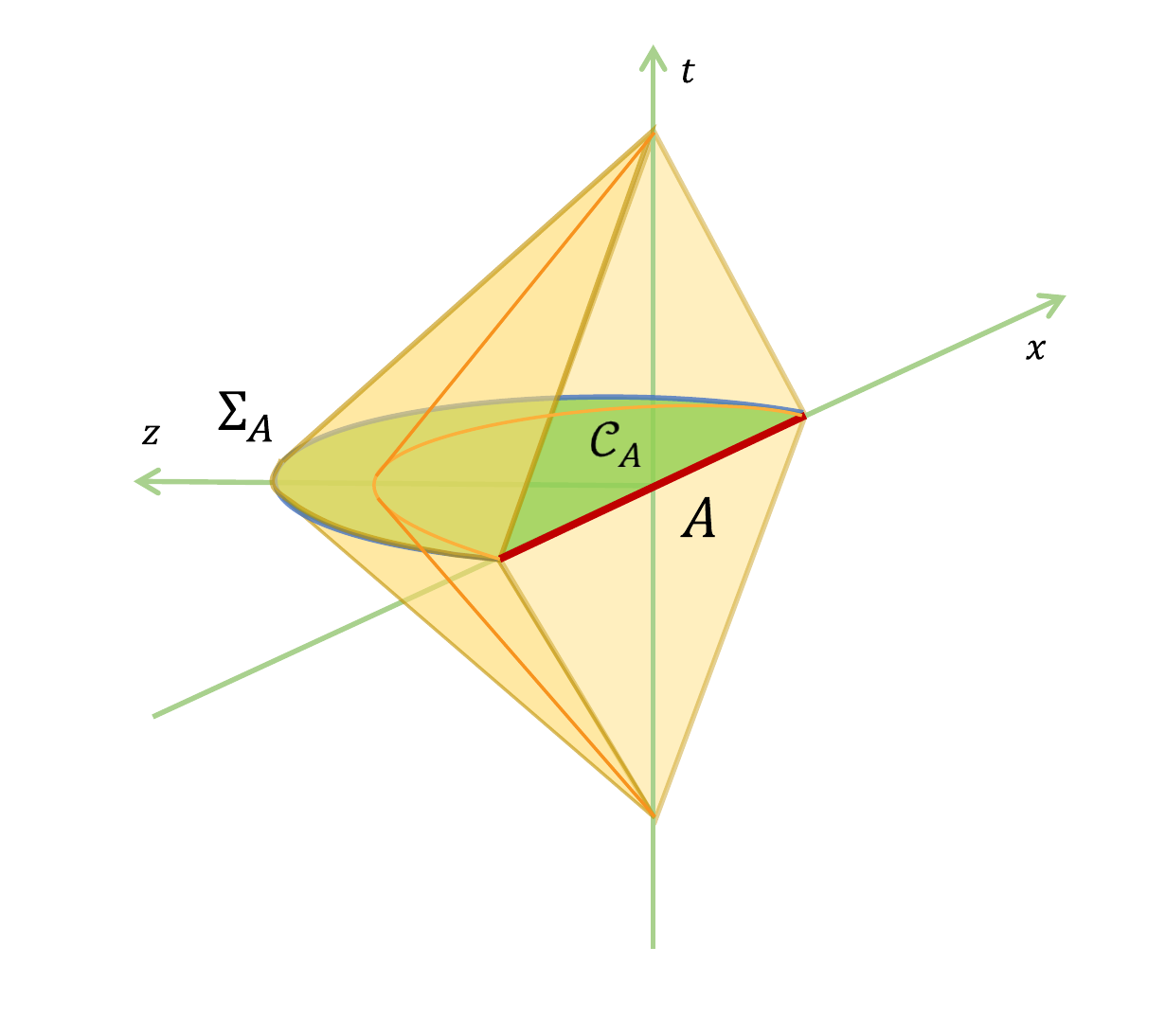}
     \caption{An illustration for Causal Wedge $\mathcal{C}_{W}(A)$ and Entanglement Wedge $\mathcal{E}_{W}(A)$.}
     \label{EW_and_CW}
 \end{figure}

A more refined notion of bulk reconstruction can be examined in the context of subregion duality \cite{Czech_2012, Headrick:2014cta}. It pertains to the following question: given a state $\psi$ and access to CFT operators supported only in a boundary subregion $A$, what is the maximal algebra of bulk local operators one can reconstruct in the effective QFT about the background geometry dual to $\psi$? Based on the algebraic structure of an effective QFT in the bulk, the answer should correspond to a particular bulk space-time wedge. The HKLL method suggests the so-called causal wedge $\mathcal{C}_W(A)$ as a candidate, which is defined to be the bulk wedge: 
\be
\mathcal{C}_W(A) = \mathcal{J}^+(D(A))\cap \mathcal{J}^-(D(A))
\ee
where $D(A)$ is the boundary Cauchy development of $A$,\footnote{We choose the definition such that $D(S)=\overline{D(S)}$ is the closure of itself.} and $\mathcal{J}^{\pm}(D(A))$ are the bulk causal future/past of $D(A)$ (See Fig \ref{EW_and_CW}).
 Reconstruction inside $\mathcal{C}(A)$ can be understood from the use of bulk-boundary propagators in the HKLL method, which also explains its constraint by causality. There is another bulk wedge naturally associated with $A$ called the entanglement wedge $\mathcal{E}_W(A)$ \cite{Czech_2012}(See Fig \ref{EW_and_CW}): 
where $\Sigma_A$ is the RT surface, or more precisely the quantum extremal surface (Q.E.S) homologous to $A$. Bulk reconstruction in $\mathcal{E}_W(A)$ was proven to exist \cite{DongXi2016} based on the JLMS proposal that identifies the bulk and boundary relative entropies \cite{Maldacena2016}. Furthermore, under the assumption that the bulk space-time satisfies the null energy condition (NEC) it was shown that the entanglement wedge is always greater or equal to the causal wedge \cite{Wall:2012,Hubeny2012,Engelhardt2015}: 
\be
\mathcal{C}_W(A) \subseteq \mathcal{E}_W(A). 
\ee
with equality holding only in special cases. Therefore, in general the causal wedge $\mathcal{C}_W(A)$ from the HKLL method underestimates the bulk reconstructibility of $A$ despite being the more intuitive guess. This implies there could exist regions in $\mathcal{E}_W(A)$ that is outside $\mathcal{C}_W(A)$, in which bulk local operators can still be reconstructible from $A$ on the boundary. Such a region is called the causal shadow of $A$ in the state $\psi$. From the causal wedge $\mathcal{C}_W(A)$ one can define the corresponding causal horizon: 
\be
\mathcal{C}_A = \dot{\mathcal{J}}^+(D(A))\cap \dot{\mathcal{J}}^-(D(A))
\ee
which is the counter-part of $\Sigma_A$ in the entanglement wedge $\mathcal{E}_W(A)$. In terms of the Q.E.S and causal horizon, the constraint $\mathcal{C}_W(A) \subseteq \mathcal{E}_W(A)$ is equivalent to the requirement that $\Sigma_A$ be everywhere deeper in the interior than $\mathcal{C}_A$ and space-like separated to it. This implies that one can place both $\mathcal{C}_A$ and $\Sigma_A$ on the same bulk Cauchy slice, e.g. for stationary or reflection symmetric states they are located on the $\tau=0$ slice. The existence of causal shadow is related to a non-empty portion of Cauchy slice between $\mathcal{C}_A$ and $\Sigma_A$.  

The mechanism for bulk reconstruction in the causal shadow of $A$, or more generally in the entanglement wedge $\mathcal{E}_W(A)$, has been a subject of keen interest because it points towards a more subtle connection between bulk and boundary physics that goes beyond apparent bulk causality. Proposals have been made regarding such reconstruction schemes, including one using eigen-modes of boundary modular flow \cite{Faulkner:2017, Faulkner:2018}; or the more information-theoretical approach using the Petz map \cite{Cotler2019, Chen2020}. One can also access the shape of the entanglement wedge from the boundary CFT  using the information theoretical Bures metric \cite{Takayanagi2019PRL, Takayanagi2020PTEP}. For the case of reconstruction using modular flow, one can understand its ability to go beyond apparent bulk causality by noticing that modular flows in general do not generate local/geometrical trajectories. Therefore they perceive a more subtle notion of boundary causality which when extended into the bulk can go beyond the notion defined by bulk-boundary propagator between local operators. As a result, one can view the existence of causal shadow as an explicit manifestation of the non-local nature of modular flow; and studying their relations thus becomes a crucial task for understanding entanglement wedge reconstruction (EWR). 

Contrary to the generic presence of causal shadow, in familiar cases where $\mathcal{E}_W(A)$ can be analytically described, e.g. spheres in conformal vacuum, they coincide with $\mathcal{C}_W(A)$ and thus give degenerate causal shadows. Not surprisingly in these cases the corresponding modular flows are local. Unfortunately, studying generic non-local modular flow and the associated causal shadow is beyond the range of our current toolbox. In this paper, we study the generation of causal shadow in the context of perturbation theory. We begin with the un-perturbed case of degenerate causal shadow, e.g. a spherical subregion $A$ in the conformal vacuum $\Omega$, where the modular flow takes local form, and consider deformations about the cases. For example, one can consider deforming the state from $\Omega$ to excited states as in \cite{Belin:2021htw}; or deforming the spherical shape of the subregion $A$, etc. These deformations will destroy the local behaviors of the modular flow which are usually the results of isometries, and thus one can expect the genesis of non-empty causal shadows in the deformed cases. Hopefully by studying the perturbation theory, we can gain some understand about the ``genetic" aspects of causal shadows.
  
\begin{figure}[h]
    \centering
    \includegraphics[scale=0.65]{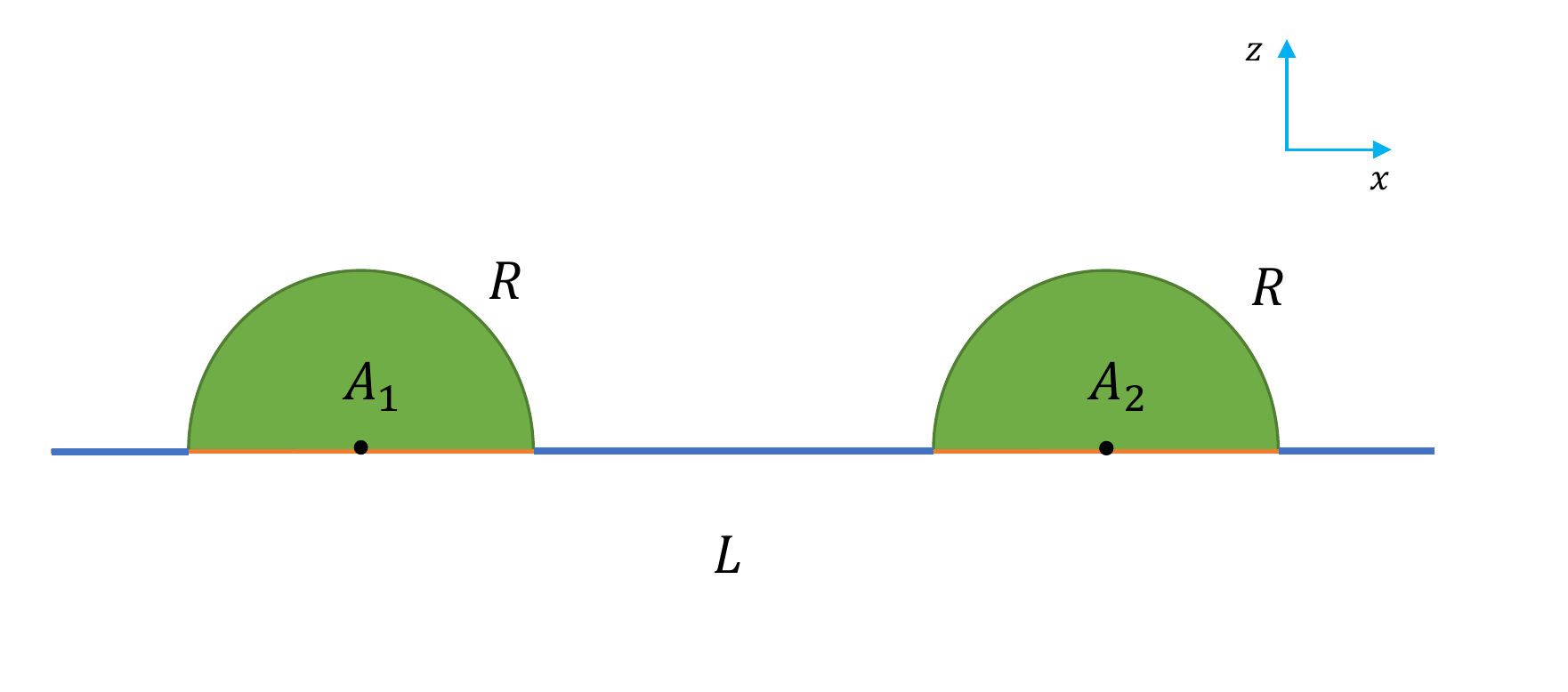}
    \caption{Two distant spheres in the conformal vacuum. The two spheres $A_{1}$ and $ A_{2} $ with radii $R_1 = R_1 = R$ are separated by a large distance $L\gg R$.}
    \label{two distant spheres}
\end{figure}

In this paper, we focus on a particular type of deformation, i.e. from correlations between two distant spheres in the conformal vacuum. More specifically, we consider the subregion as consisting of two spheres $A_{1}$ and $A_{2}$ of radius $R_1 = R_2 = R$ separated by a large distance $L\gg R$ (See Fig \ref{two distant spheres}).

 One can view this as a deformation to the $L=\infty$ limit, where the reduced density matrix factorizes $\rho_{A_{1}\cup A_{2}} = \rho_{A_{1}} \otimes \rho_{A_{2}}$. For finite $L\gg R$, the effects on each spherical component due to the other one is small, and one can do perturbation theory in $R/L$. To clarify further, let us understand the effect of such ``correlational" deformation from both the boundary and bulk point of views. From the boundary perspective, the modular flow associated with each sphere is local, but their corresponding isometries are not compatible; therefore once we include both spheres as the subregion, the modular flow cease to be local in each component due to the correlation. From the bulk perspective, in the $L\gg R$ regime the classical minimal (RT) surface is simply the union of the minimal surfaces from both components, so is the causal horizon; on the other hand, the Q.E.S receives a correction relative to the causal horizon due to the mutual information of bulk QFT between the two components, and therefore a causal shadow emerges from such quantum corrections in $G_N$. As result, the causal shadow in our set-up admits a double-expansion: boundary ``OPE" in $R/L$ and bulk quantum corrections $G_N$; our goal is to study the leading order result for both expansions. One nice feature of the set-up is that the causal horizon, which is usually more difficult to capture in generic cases, remains intact under this deformation; this makes it easier to study properties of causal shadows, and is the main reason for our choice. 

This paper is organized as follows. In section (\ref{sec:degenerate}) we revisit known cases with degenerate causal shadow, the purpose is to relate local behavior of modular flow to the degeneracy of causal shadows in a more general context; in section (\ref{sec:bulk_MI}) we compute the linear shape dependence of bulk mutual information between two entanglement wedges, which is the driving ingredient giving rise to a non-degenerate causal shadow; in section (\ref{sec:main}) we obtain the geometry of the causal shadow by solving the bulk Q.E.S equation to the leading order in both $G_N$ and $R/L$; in section (\ref{sec:comments}) we discuss the properties and implications of perturbative causal shadows from the general perspective of boundary CFTs, and explore the possibility of extracting a positivity condition that characterizes the causal shadows; in section (\ref{sec:discuss}) we summarize the paper and propose some future works. 

\vspace{5mm}

\section{From local modular flow to degenerate causal shadows}\label{sec:degenerate}
Before embarking on the perturbative study of causal shadows, we begin in this section by revisiting the degenerate case $\mathcal{C}_W(A) = \mathcal{E}_W(A)$. Examples of such cases include the Rindler half-plane in relativistic QFT vacuum and spherical region in CFT vacuum. In each example one can check explicitly the coincidence between the entanglement wedge and causal wedge as well as the local nature of associated modular flow. In order to further understand the relation between these two phenomena, in this section we will construct an argument that starts from the local behavior of modular flow and leads to the degeneracy of causal shadow. Due to the potential subtleties in dealing with the most general cases, e.g. issues related to analytic continuation between Euclidean and Lorentzian geometries, we do not pretend to provide a rigorous and general proof for the implication relations between the two phenomena. Whenever needed we will make convenient assumptions that are true in the known examples -- the point is rather to identify the essential logic that control the degeneracy of causal shadow in these cases. 

\begin{figure}[h]
     \centering
     \includegraphics[scale=0.50]{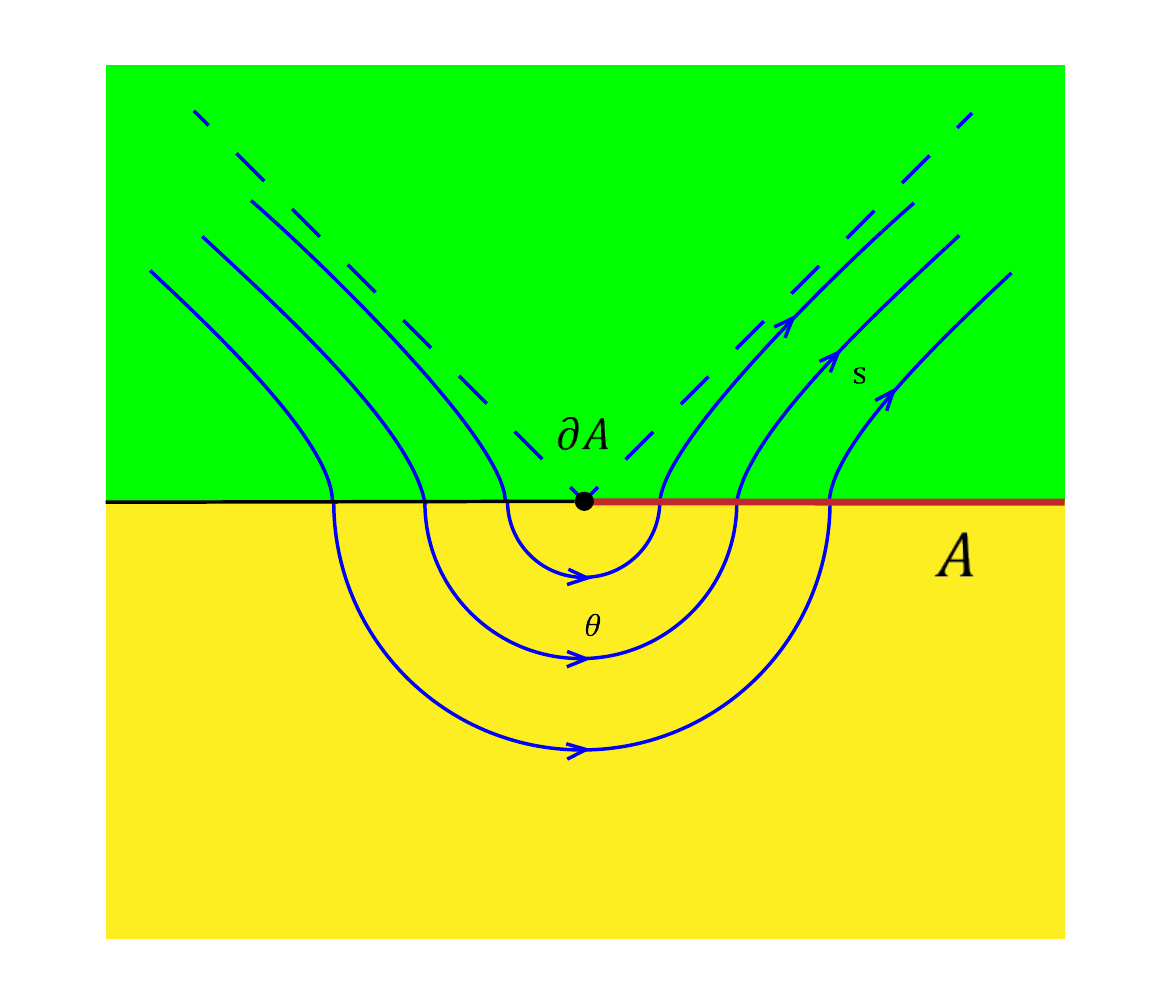}
     \caption{Analytic continuation between local Lorentzian (top) and Euclidean (below) modular flows.}
     \label{fig:E_L}
 \end{figure}

Before proceeding with the argument, let us first set up the context of our discussions. To be more generic, we allow the boundary CFT to be defined with a background metric $G$ and possible sources $J$ turned on. For simplicity we only consider the vacuum state $\Omega$ and a simply connected subregion $A$, out of which we can construct a reduced density matrix $\rho_A = \text{tr}_{\bar{A}}|\Omega\rangle \langle \Omega|$. \footnote{Strictly speaking, in QFTs the tensor-product structure $\mathcal{H} = \mathcal{H}_A \otimes \mathcal{H}_{\bar{A}}$ of the Hilbert space and thus the existence of reduced density matrix is problematic. There exists more rigorous ways to construct the various objects in our discussion using the algebraic framework \cite{Haag_1996}. In this paper, we shall take the less rigorous approach using the reduced density matrix. } We assume that $\Omega$ is well-defined on $G$ and admits a Euclidean path-integral definition for its Hartle-Hawking wave-functional, so that the reduced density matrix elements can be written schematically as:
\be\label{eq:RDM}
\left[\rho_A\right]^\alpha_\beta  = \int \left[\mathcal{D} \Phi\right]^{\Phi^+(A)=\alpha}_{\Phi^{-}(A)=\beta} \;e^{-I_E(G,J,\Phi)}
\ee
where $\Phi$ denotes the boundary quantum fields collectively. From the reduced density matrix one can define the corresponding modular Hamiltonian: 
\be
\rho_A = e^{-2\pi K_A} 
\ee
and use it to define the modular flow $\mathcal{O}\to \mathcal{O}^{(s)}$ by conjugation on operators: 
\be
 \mathcal{O}^{(s)} = e^{is K_A } \mathcal{O} e^{-isK_A}
\ee
To appreciate the choice of terminology, we recall that when the density matrix $\rho$ takes the special form of canonical ensemble, the modular Hamiltonian becomes the physical Hamiltonian and modular flow is simply the time evolution in Heisenberg picture. When the modular flow maps a local operator to another local operator, i.e.
\be\label{eq:local}
\mathcal{O}^{(s)}(x) \propto \mathcal{O}(x(s)),\;\; x \in A, x(s) \in \mathds{R}^{d-1,1}
\ee
then we refer to such a modular flow as being local. The locus $\lbrace x(s),s\in\mathds{R} \rbrace$ is called the modular trajectory through $x$. To better relate to the Euclidean path-integral definition of $\rho_A$, we assume that there exists an analytic continuation of (\ref{eq:local}) via wick-rotating $s\to i\theta$:
\be\label{eq:E_local}
\mathcal{O}^{(i\theta)}(x) \propto \mathcal{O}(x(i\theta)),\;\; x\in A, x(i\theta) \in \mathds{R}^{d}
\ee
that moves from the Lorentzian to the Euclidean section, see Fig (\ref{fig:E_L}). There is a generalized KMS condition for the modular flowed operator, which dictates that the Euclidean modular trajectories form closed loops with period $2\pi$: 
\be
x(i\theta) = x(i\theta+2\pi i), \;\;\forall x \in A  
\ee

We now argue that these properties imply that the causal shadow is degenerate, i.e. $\mathcal{C}_W(A) = \mathcal{E}_W(A)$. Recall that the Q.E.S is defined by the following extremization property of the generalized entropy $S_{gen}$: 
\be\label{eq:QES}
\Sigma_A = \text{Ext}_{\Sigma'} \left\lbrace S_{gen}\left(\Sigma'\right)=\frac{\text{Area}\left(\Sigma'\right)}{4G_N}+ S_{bulk}\left(a'\right), \partial \Sigma' = \partial A, \partial a' = \Sigma'\cup A\right\rbrace 
\ee 
where $S_{bulk}(a)$ is the entanglement entropy of the bulk effective QFT on the bulk subregion $a$. The extremization is over bulk surfaces homologous to $A$, while recent developments \cite{Almheiri:2019} have suggested it can also include additional closed regions, i.e. ``islands". We do not consider possible issues with islands in this paper. Following the decomposition of $S_{gen}$ into the classical contribution and quantum corrections, our argument also consists of two parts. The first part concerns the RT surfaces, we argue that they coincide with causal horizons; the second part concerns quantum corrections, and we argue that they do not shift the Q.E.S relative to RT surfaces.  

\subsection{Classical analysis}\label{subsec:classic}
We first argue that in the case of local modular flow, the classical RT surface $\tilde{\Sigma}_A$ coincides with the causal horizon $\mathcal{C}_A$. In its essence, this is a statement regarding both the boundary and bulk geometries. So we proceed by analyzing the implications of the local nature of modular flow defined in (\ref{eq:E_local}) about the underlying geometries.

\subsubsection{Group of conformal isometries}
We begin with the boundary CFT. Local modular flows are highly non-generic, because it imposes strong geometrical constraint on the defining state and subregion. For example, requiring (\ref{eq:E_local}) to hold for all $\mathcal{O}$ implies that there is an $U(1)$ group of isometries associated with the Euclidean path-integral (\ref{eq:RDM}). In particular, this means that the background metric $G$ and external sources $J$ are all invariant under the action: 
\be
G\circ U(\theta) = G,\;\; J\circ U(\theta) = J,\;\;0\leq \theta \leq 2\pi
\ee
In addition, the entangling surface $\partial A$ is fixed under the $U(1)$ isometry group actions. In the case of CFTs, the constraint can be relaxed to accommodate conformal isometries, i.e.
\be 
G\circ U(\theta)(x) = \Lambda_\theta(x)^{-2} G(x)
\ee 
in which case there will be an additional factor of $\Lambda_\theta(x)^{ -\Delta }$  on the RHS of
(\ref{eq:E_local}). 

In the context of AdS/CFT, these (conformal) isometries are part of the asymptotic boundary conditions. We assume that these symmetries are not ``spontaneously broken" by the dominant bulk saddle dual to the state $\Omega$, i.e. there exist bulk extensions of the $U(1)$ group of boundary isometries or conformal isometries -- in both cases their bulk extensions are ordinary isometries of the bulk metric $g_{\mu\nu}$. Furthermore, the classical RT surface $\tilde{\Sigma}_A$ is fixed under the bulk isometries, they are simply the bulk extension of the boundary fixed-surface $\partial A$. \footnote{Notice we placed a tilde in the RT surface $\tilde{\Sigma}_A$ to distinguish from the quantum extremal surface $\Sigma_A$.} This is a non-trivial assumption we make about bulk dynamics which is valid in the known examples we consider. One can then expand the metric about $\tilde{\Sigma}_A$ in Gaussian-normal coordinates as: 
\be \label{eq:GN_E}
ds^2 = dr^2+g_{\theta\theta}(r,y) d\theta^2+ g_{ij}(r,y) dy^j dy^j,\;\;i,j = 2,...,d+1
\ee
where $r$ is the proper length along radial geodesics orthogonal to $\tilde{\Sigma}_A$, and smoothness of the Euclidean geometry implies that $g_{\theta\theta}(r,x)=r^2\left(1+\mathcal{O}(r)\right)$ as $r\to 0$. The bulk isometries correspond to rotations about $\tilde{\Sigma}_A$: $\theta \to \theta+\text{const}$. The fixed-surface $\tilde{\Sigma}_A$ is automatically a minimal surface, in fact it is a totally geodesic surface. A totally geodesic sub-manifold is special in that its extrinsic curvature tensors vanish identically $K^{a}_{ij}=0$ while a minimal surface only requires their traces to vanish: $\text{tr}K^{a}=0$.

Next we assume the analytic continuation $\theta \to is$ can also be extended into the bulk saddle point geometry, which moves from the Euclidean bulk saddle into the Lorentzian section with metric: 
\be\label{eq:GN_L}
ds^2 = dr^2-g_{\theta\theta}(r,y) ds^2+ g_{ij}(r,y) dy^j dy^j,\;\;i,j = 2,...,d+1
\ee 
The bulk isometries guarantee that such a bulk Lorentzian section is well-defined, on which the entanglement wedge is defined. Now we have a $\mathds{R}$ group of (conformal) isometries: $s\to s+\text{const}$ from the (Lorentzian) boundary extended into the bulk. In particular, this implies that the local modular trajectories $x(s)$ are also extended into the bulk. We assume the Hartle-Hawking state $\Omega$ has a reflection symmetry under $\tau\to -\tau$ which again is respected by the bulk Euclidean saddle, then the modular trajectories for real $s$ are time-like curves in space-time. This follows in two steps. Firstly, the reflection symmetry in Lorentzian $t\to -t$ implies that the boundary/bulk modular trajectories are orthogonal to $A$/$a$, which are part of the $t=0$ boundary/bulk Cauchy slice. This is because they are integral curves of (conformal) isometries and so must also be reflection symmetric under $t\to -t$. Secondly, being time-like at the intersections with $A$/$a$ and being integral curves of (conformal) isometries, the modular trajectories must remain time-like throughout the entire journey.  

\subsubsection{Caustics structures on null geodesic congruences}
On the boundary, as we take the limit $x\to \partial A$ for $x\in A$, the modular trajectories $x(s)$ approach future/past directed inward null geodesics from $\partial A$ orthogonally for positive/negative $s$, as can be inferred from the $r\to 0$ limit of (\ref{eq:GN_L}). In other words, the inward orthogonal null geodesic congruences $\mathcal{N}^{\pm}(\partial A)$ from $\partial A$ coincide with limiting class of modular trajectories: 
\be\label{eq:bdry_null}
\mathcal{N}^{\pm}(\partial A) = \left\lbrace \lim_{x\to x'} x(s), x\in A, s\in \mathds{R}^{\pm}, x'\in \partial A \right\rbrace 
\ee
Similarly for the inward orthogonal null geodesic congruences $\mathcal{N}^{\pm}(\tilde{\Sigma}_A)$ in the bulk from $\tilde{\Sigma}_A\subset \partial a$, we have that: 
\be\label{eq:bulk_null}
\mathcal{N}^{\pm}(\tilde{\Sigma}_A) = \left\lbrace \lim_{x\to x'} x(s), s\in \mathds{R}^{\pm}, x\in a, x'\in \tilde{\Sigma}_A \right\rbrace 
\ee
The distributions of the modular parameter $s$ along the null geodesic congruences become singular in the $r\to 0$ limits of (\ref{eq:bdry_null},\;\ref{eq:bulk_null}); a more natural parametrization is to foliate them into slices of constant affine parameters $(u,v)$. In order to do this, the $r\to 0$ and $s\to \pm \infty$ limits need to be taken simultaneously so as to keep the affine parameters $u=r e^{s}$ on $\mathcal{N}^{+}$, or $v=re^{-s}$ on $\mathcal{N}^{-}$ fixed. 

\begin{figure}[h]
     \centering
     \includegraphics[scale=0.50]{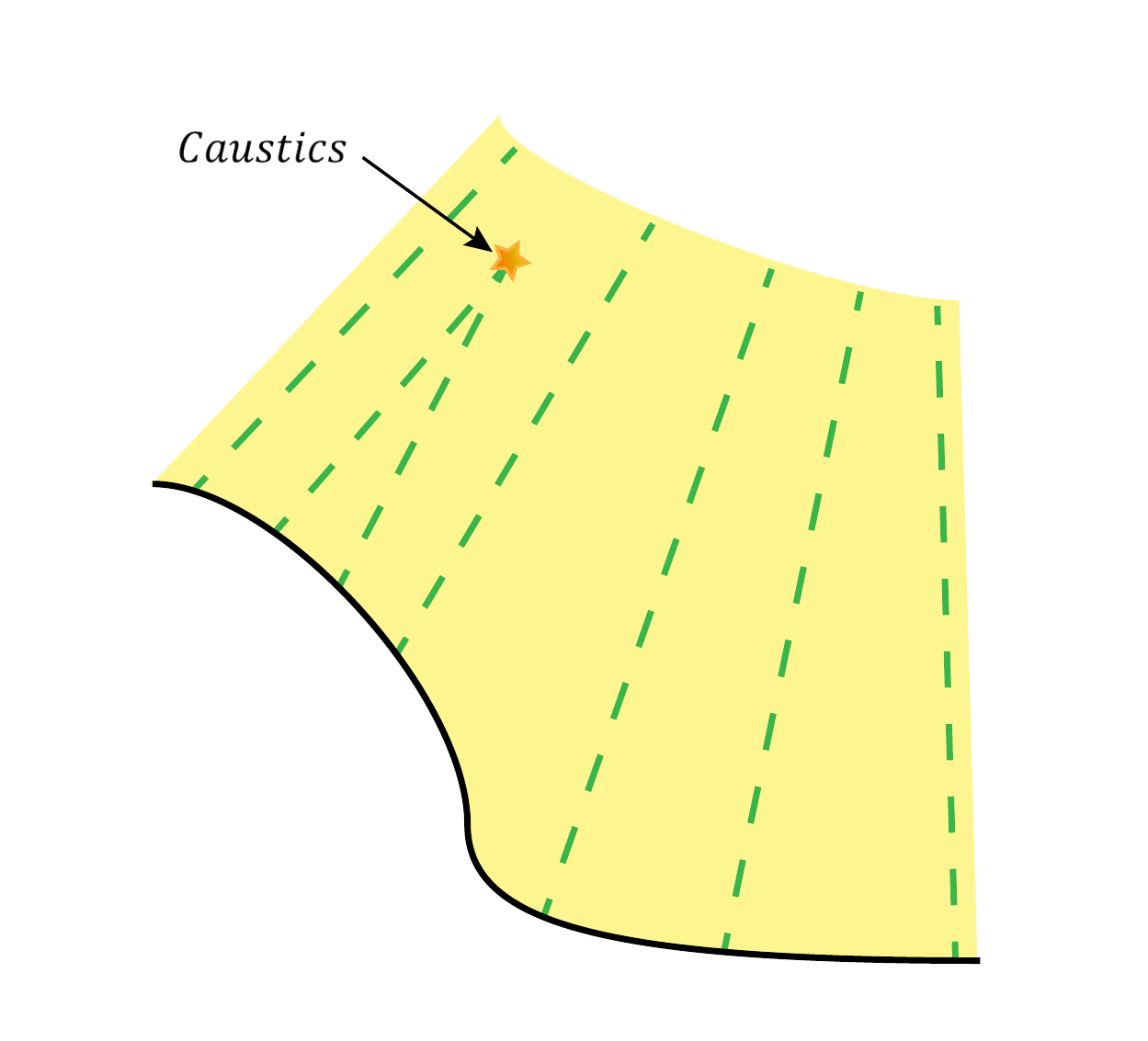}
     \caption{Caustics are formed on a null geodesic congruence when nearby null generators intersect.}
     \label{fig:caustics}
 \end{figure}

The above characterization leads to the following crucial property regarding these null congruences: they do not develop caustics except at singularities of the Lorentzian conformal factors, which can only be on the boundary. Roughly speaking, caustics are space-time incidents where nearby null geodesics in the congruence cross one another, see Fig (\ref{fig:caustics}). For simplicity we focus on the $\mathcal{N}^{+}$ components. We first discuss the bulk congruence $\mathcal{N}^{+}(\tilde{\Sigma}_A)$. Since the bulk modular trajectories are integral curves of isometries, therefore proper distances $dx(s)=|x'(s)-x(s)|$ between nearby points at $x$ are preserved under $s$ evolution:
\be
dx(s)=dx(0),\;\; x,x'\in a
\ee 
Taking the appropriate $x,x'\to y,y'\in \tilde{\Sigma}_A$ limits while keeping the affine parameter $u$ fixed, we have that orthogonal proper distances $dy(u)$ between nearby null geodesics on $\mathcal{N}^{+}(\tilde{\Sigma}_A)$ are preserved under shifts in $u$, and thus no bulk caustics can develop on these null geodesic congruences from $\tilde{\Sigma}_A$ -- except the limit when they reach the asymptotic boundary. So next we discuss the boundary congruence $\mathcal{N}^{+}(\partial A)$. In this case, isometries can be conformal, so near the fixed points $\partial A$ one can write the boundary Euclidean and Lorentzian metrics as:
\bea\label{eq:GNC_bdry}
&& ds_E^2 = \Lambda^2_E(\theta, r, y) \Big(dr^2+\tilde{G}_{\theta\theta}(r,y) d\theta^2+ \tilde{G}_{ij}(r,y) dy^j dy^j\Big)\nonumber\\
&& ds_L^2 = \Lambda^2_L(s, r, y) \Big(dr^2-\tilde{G}_{\theta\theta}(r,y) ds^2+ \tilde{G}_{ij}(r,y) dy^j dy^j\Big), i,j = 2,...,d
\eea
where we assume the analytic continuation between Lorentzian and Euclidean conformal factors $\Lambda_L(s,r,y)=\Lambda_E(is,r,y)$ is well-defined. As a result of the conformal isometries, the orthogonal proper distances $dy(u)$ between nearby null geodesics do change with the affine parameters $u$ along $\mathcal{N}^{+}(\partial A)$, but the ratios are controlled by the conformal factors: 
\be 
dy(u) = \left[\frac{\Lambda(u,y)}{\Lambda(0,y)}\right]dy(0),\;\;\Lambda(u,y) = \lim_{r\to 0}\left\lbrace \Lambda_L(s,r,y),\;r e^{s}=u\right\rbrace,\;\;y\in \partial A
\ee
Therefore away from singularities of $\Lambda(u,y)$ or the similarly defined $\Lambda(v,y)$, no caustics can develop along $\mathcal{N}^{\pm}(\partial A)$. For example, for $A$ being the spherical region in the conformal vacuum, the tips of its null-cone are caustics of this nature. From now on we consider the conformal frame in which $A$ is bounded, so that caustics are guaranteed to arise. This will avoid the issue of identifying infinities. 

We shall assume that the null-cone tips are good representatives of the caustics that can arise at the level of generality we wish to consider regarding local modular flow. In more precise terms, we assume that  caustics can only appear as space-time points where all orthogonal null geodesics on $\mathcal{N}^{\pm}(\partial A)$ converge at the same point -- in fact together with other modular trajectories $\left\lbrace x(s), s\in \mathds{R}, x\in A\right\rbrace$. Admittedly this is a crucial and strong assumption, so we should provide some heuristics for its plausibility. Using the metric ansatz (\ref{eq:GNC_bdry}) for $G$, we can write its scalar curvature $\mathcal{R}_G$ as: 
\be\label{eq:curvatures}
\mathcal{R}_G = e^{2f}\Big(\mathcal{R}_{\tilde{G}} +  2(d-1) \nabla^2_{\tilde{G}}\; f -(d-2)(d-1) \big|\nabla_{\tilde{G}} f\big|^2\Big),\;\;\Lambda_L=e^{-f}
\ee 
where $\tilde{G}$ is the rescaled metric: 
\be
ds^2 = dr^2-\tilde{G}_{\theta\theta}(r,y) ds^2+ \tilde{G}_{ij}(r,y) dy^j dy^j
\ee
With both $\tilde{G}$ and $G$ fixed, (\ref{eq:curvatures}) can be viewed as a complicated non-linear PDE for $f$:
\be
 2(1-d) \nabla^2_{\tilde{G}}\; f +(d-2)(d-1) \big|\nabla_{\tilde{G}} f\big|^2 -\mathcal{R}_{\tilde{G}}+e^{-2f} \mathcal{R}_G =0
\ee 
As discussed before, the caustics are where the ``dilaton" diverges: $f\to \infty$. The structure of such caustics is controlled by the following constraints:
\begin{enumerate} 
\item The scalar curvature $\mathcal{R}_{\tilde{G}}$ is smooth and independent of $s$.
\item The physical metric $G$ is also smooth.
\item In particular, $\mathcal{R}_{G}$ remains finite at the caustics where $f\to \infty$.
\end{enumerate}
As a result one can neglect terms proportional to $\mathcal{R}_G$ and $\mathcal{R}_{\tilde{G}}$ near the caustics, where the condition (\ref{eq:wave}) becomes approximately a quasi-linear wave equation in $(s,x)$ coordinates with Dirichlet boundary condition imposed at $\partial A$:
\be\label{eq:wave}
\nabla^2_{\tilde{G}} f(s,x) \approx \left(\frac{d-2}{2}\right)\left|\nabla_{\tilde{G}}f(s,x)\right|^2,\;\;f\to \infty
\ee
where  $s\in \mathds{R}$ is the time coordinate and $x=(r,y)\in A$ represent spatial coordinates. The non-linear term $\left|\nabla_G f\right|^2$ shifted to the RHS can be viewed effectively as a dynamical source for the wave equation. One likely scenario is that $f(s,x)\to \infty$ as $s\to \pm \infty$ collectively for all $x\in A$, which is the case for a 
vacuum sphere in CFTs. More generally we may expect any local diverging behaviors of $f$ to drive the wave while also propagate spatially throughout $A$. As a result for bounded and simply connected $A$, the divergences of $f$, if present, cover the whole spatial domain: i.e. the conformal factor $\Lambda_L \to 0$ along every modular curve $\lbrace x(s), s\in \mathds{R}\rbrace$ for $x\in A$. This means that the physical volumes of spatial-slices for the ``modular-flow wedge": 
\be
M_A=\left\lbrace x(s),\;s\in \mathds{R},x\in A\right\rbrace
\ee 
shrinks to zero as we approach the caustics, i.e. they become tips of the null congruences $\mathcal{N}^{\pm}(\partial A)=\partial M_A$. This is indeed a crude heuristics regarding the plausibility of our assumption: tip of null cone is the prototype of caustics when modular flow is local. Strictly speaking we could not rule out other possibilities e.g. that they form a fractal structure also with zero-volume. A more careful analysis would involve a systematic treatment of the non-linear PDE (\ref{eq:wave}) that is beyond the scope of this work, we leave this for future investigations. 

\subsubsection{Coincidence between RT surface and causal horizon}
Proceeding with the assumption, it then implies that the null geodesic congruences $\mathcal{N}^{\pm}(\partial A)$ coincide with the Cauchy horizon $\mathcal{H}^{\pm}(A)$. Recall that the Cauchy horizon of an achronal set $S$ is the boundary of its Cauchy development: 
\be
\mathcal{H}^{\pm}(S) = \dot{D}^{\pm}(S)
\ee
where $\pm$ denotes the future/past component. We then recall the following property of Cauchy horizon $\mathcal{H}^{\pm}(S)$ starting from $\partial S$: they coincide with the null geodesic congruences $\mathcal{N}^{\pm}(\partial S)$ all the way until encountering caustics \cite{hawking_ellis_1973}. Beyond caustics the null geodesic exits the Cauchy horizon $\mathcal{H}^{\pm}(S)$  -- if $\mathcal{H}^{\pm}(S)$ do not happen to end there, and enters the interior of $D^{\pm}(S)$. Under our assumption, the caustics on $\mathcal{N}^{\pm}(\partial A)$ are null-cone tips where all orthogonal null geodesics converge; it is easy to see that they are also the future/past end-points of the Cauchy horizons $\mathcal{H}^{\pm}(A)$. Therefore their coincidence is global, i.e.
\be 
\mathcal{N}^{\pm}(\partial A) = \mathcal{H}^{\pm}(A) = \dot{D}^{\pm}(A)
\ee 

\begin{figure}[h]
     \centering
     \includegraphics[scale=0.75]{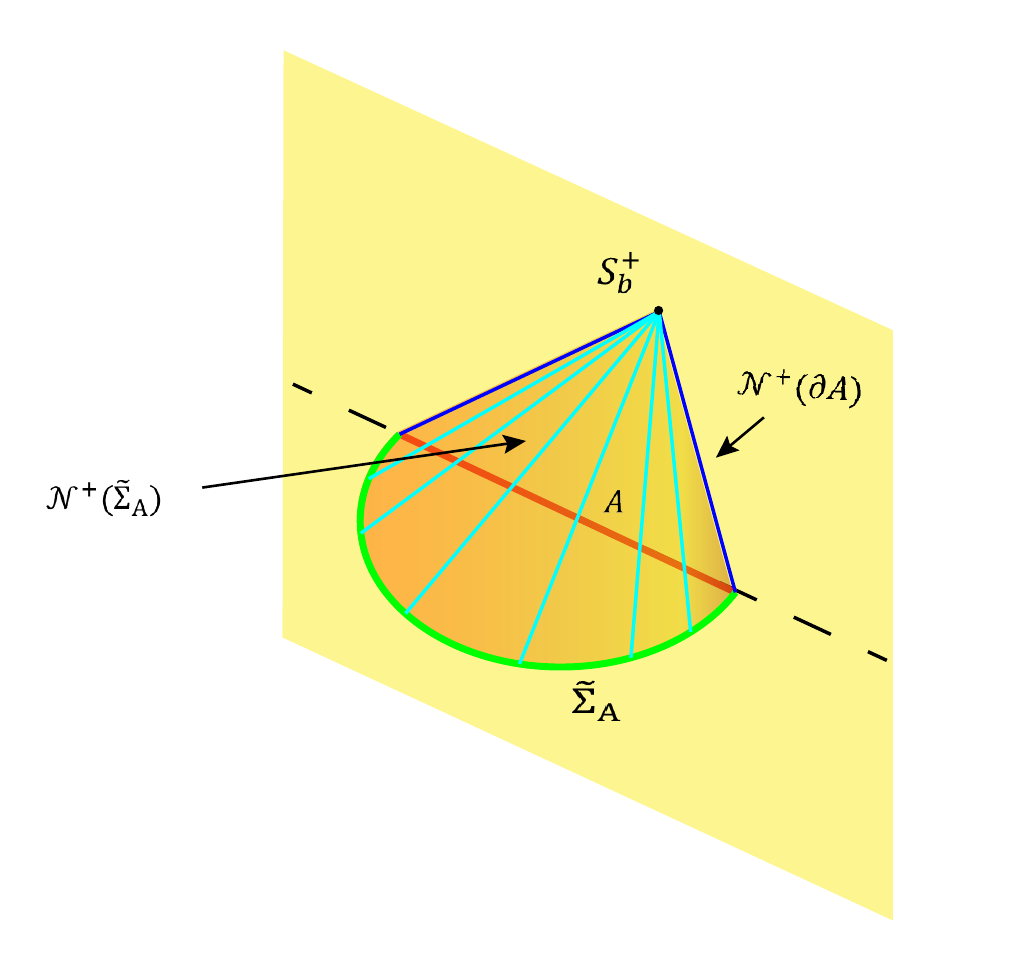}
     \caption{We are only showing the future components. In the absence of the bulk caustics on it, $\mathcal{N}^+\left(\tilde{\Sigma}_A\right)$ intersects with the asymptotic boundary at $\mathcal{N}^+(\partial A)\cup S^+_b$, with $S^+_b=S^+_A$ in the form of a null-cone tip in our cases.}
     \label{fig:null_cone}
 \end{figure}

Finally we move into the bulk and argue the coincidence between RT surface and causal horizon: $\tilde{\Sigma}_A = \mathcal{C}_A$. Recall that the causal horizon is defined by the intersection of two causal boundaries in the bulk that are of the event-horizon type: 
\be
\mathcal{C}_A = \dot{\mathcal{J}}^+\left(D(A)\right)\cap  \dot{\mathcal{J}}^-\left(D(A)\right)
\ee
while the two null geodesic congruences intersect at the RT surface: 
\be
\tilde{\Sigma}_A = \mathcal{N}^+(\tilde{\Sigma}_A) \cap  \mathcal{N}^-(\tilde{\Sigma}_A)
\ee
It is therefore sufficient to show that:
\be 
\mathcal{N}^{\pm}(\tilde{\Sigma}_A) = \dot{\mathcal{J}}^{\mp}\left(D(A)\right)
\ee 
between $\tilde{\Sigma}_A$ and the asymptotic boundary. Analogous to the case of Cauchy horizons, the event-horizon type boundaries $\dot{\mathcal{J}}^{\pm}\left(D(A)\right)$ coincide with the bulk orthogonal null geodesic congruences $\mathcal{N}^{\pm}(S^\mp_A)$ from $S^{\pm}_A$ until encountering caustics, where $S^{\pm}_A$ are the future/past boundaries of the Cauchy horizons $\mathcal{H}^{\pm}(A)$. In general $S^{\pm}_A$ are achronal sets, but in our cases they are simply the future/past null-cone tips. On the other hand, we have shown that there is no bulk caustics on $\mathcal{N}^{\pm}(\tilde{\Sigma}_A)$, thus they are smooth null surfaces in the bulk. As a result their intersections with the asymptotic boundary, denoted $\partial \mathcal{N}^{\pm}(\tilde{\Sigma}_A)$, is connected.  Furthermore, the RT surface intersects the asymptotic boundary at $\partial \tilde{\Sigma}_A = \partial A$, therefore we have that:
\be
\partial \mathcal{N}^{\pm}(\tilde{\Sigma}_A) = S^{\pm}_b\cup \mathcal{N}^{\pm}(\partial A) = S^{\pm}_b \cup \mathcal{H}^{\pm}(A)
\ee
where $S^{\pm}_b$ denote the intersections on the asymptotic boundary of all the bulk orthogonal null geodesics on $\mathcal{N}^{\pm}(\tilde{\Sigma}_A)$, see Fig (\ref{fig:null_cone}). 
Hence $S^{\pm}_b \cup \mathcal{H}^{\pm}(A)$ is connected, and because $\mathcal{H}^{\pm}(A)$ is already a connected null cone with the future/past tip $S^{\pm}_A$, we must have that $S^{\pm}_b = S^{\pm}_A$. We conclude that all orthogonal null geodesics on $\mathcal{N}^{\pm}(\tilde{\Sigma}_A)$ converge to the asymptotic boundary at $S^{\pm}_A$. Reversing the direction, $\mathcal{N}^{\pm}(\tilde{\Sigma}_A)$ can be viewed as null geodesic congruences from $S^{\pm}_A$ to $\tilde{\Sigma}_A$ that contain no caustics, and thus coincide with $\dot{\mathcal{J}}^{\mp}\left(D(A)\right)$ in between. This completes our argument for the coincidence between RT surface and causal horizon.

\subsection{Quantum corrections}\label{subsec:quantum}
At small but finite $G_N$, the entanglement wedge $\mathcal{E}_W(A)$ is instead defined by the Q.E.S which extremizes the generalized entropy $S_{gen}$ in (\ref{eq:QES}) rather than the geometric area like RT surface does. In general, this is a perturbative modification to the extremality condition that will result in a correction in the location of Q.E.S relative to the classical RT surface. To argue that $\mathcal{C}_W(A) = \mathcal{E}_W(A)$, or the coincidence between Q.E.S and causal horizon, we should argue in addition to subsection (\ref{subsec:classic}) that the Q.E.S coincides with the classical RT surface, i.e. $\Sigma_A = \tilde{\Sigma}_A$. This amounts to saying that the classical RT surface, apart from being a minimal surface, also extremizes the generalized entropy: 
\be
\left(\frac{\partial S_{gen}(\Sigma')}{\delta \Sigma'}\right)_{\tilde{\Sigma}_A} = \frac{1}{4G_N}\left(\frac{\delta \text{Area}(\Sigma')}{\delta \Sigma'} \right)_{\tilde{\Sigma}_A}+ \left(\frac{\delta S_{bulk}(a')}{\delta \Sigma'}\right)_{\tilde{\Sigma}_A} =0,\;\;\partial a'=\Sigma'\cup A
\ee
The classical RT surface is defined by the vanishing of the first area term, so its coincidence with the Q.E.S is dictated by satisfying: 
\be
 \left(\frac{\delta S_{bulk}(a')}{\delta \Sigma'}\right)_{\tilde{\Sigma}_A} =0
\ee 
In order to show this, we need to work with the shape-perturbation theory of bulk entanglement entropy defined on the bulk saddle $\mathcal{B}$ as background, compute the linear response of entanglement entropy under infinitesimal shape deformations $\tilde{\Sigma}_A \to \Sigma'=\tilde{\Sigma}_A+ \lambda\;\delta \Sigma,\lambda \ll 1$: 
\be
S_{bulk}(a')-S_{bulk}(a) = \lambda\; \delta S_{bulk}(a) + \mathcal{O}\left(\lambda^2\right)
\ee
and show that the linear response $\delta S_{bulk}(a)=0$ for deformations confined in the bulk, i.e. do not change the boundary subregion $A$. 

\subsubsection{Shape perturbation theory and entanglement first law}
\begin{figure}[h]
    \centering
    \includegraphics[scale=0.50]{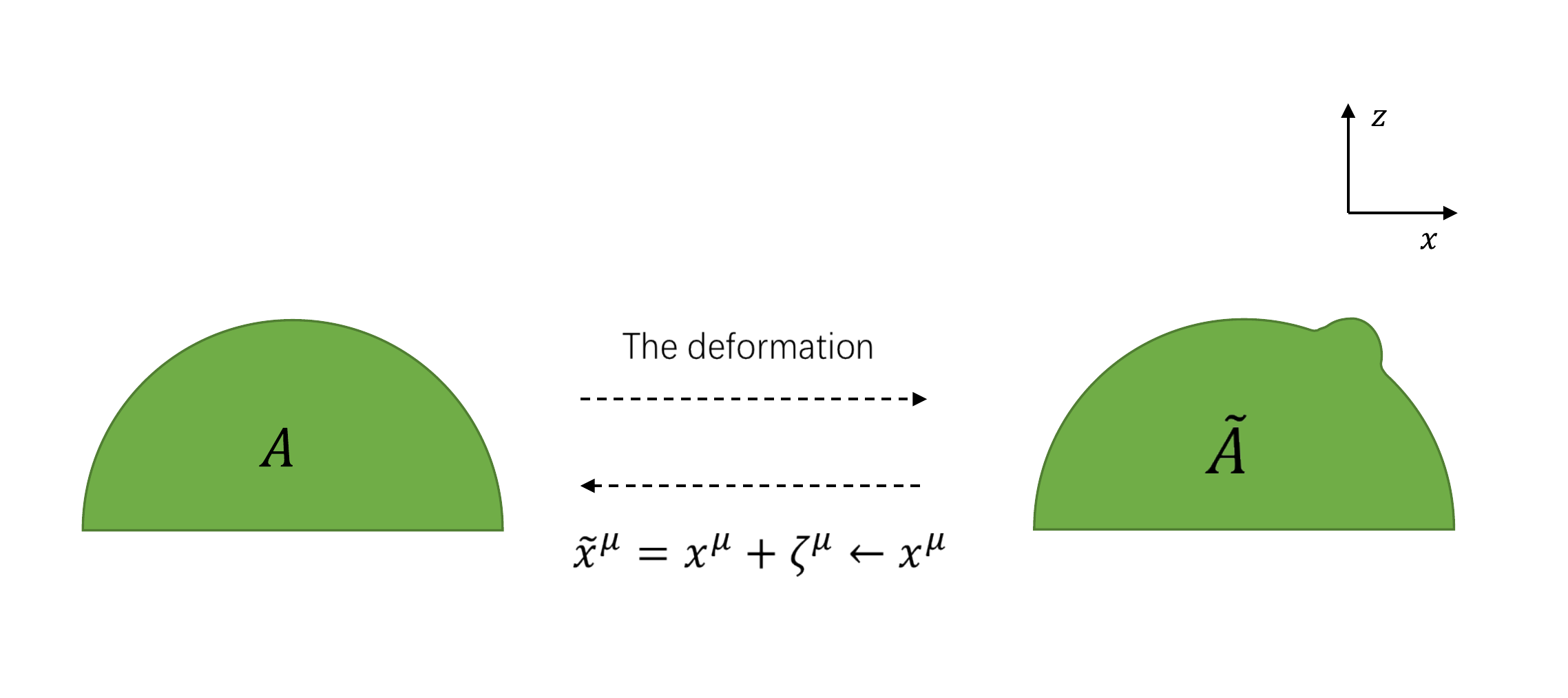}
    \caption{Relating shape deformation to metric deformation via a corresponding diffeomorphism.}
    \label{the deformation of subregion-setup}
\end{figure}

Let us quickly recall the basic ingredients involved in the computation, which also applies to latter computations. More details can be found in \cite{Rosenhaus2014Dec, Rosenhaus2015Feb,Mezei2015Feb, Mezei2015Feb02,Faulkner2016Apr}. At the leading order in $G_N$ relevant for the bulk QFT, we can assume the quantum state to be in the ground-state whose wave-functional admits a Hartle-Hawking type Euclidean path-integral in the lower-half-plane of the bulk saddle $\mathcal{B}$. As a result, the reduced density matrix elements on a bulk subregion $a$ can be written in terms of the Euclidean path-integrals with branch-cuts on $a$ similar to (\ref{eq:RDM}):
\be
\langle \alpha|\rho_a(g,\;j) |\beta\rangle =  \int_{\phi\left(a^{-}\right)=\beta}^{\phi\left(a^{+}\right)=\alpha}[\mathcal{D} \phi] e^{-I^{bulk}_{E}\left(g,\; j,\;\phi \right)}
\ee
where $\phi$ denotes the bulk quantum fields collectively, and $(g,j)$ represent the bulk metric and sources of the bulk saddle $\mathcal{B}$. Now we perform an infinitesimal shape deformation of the entangling region $a \rightarrow \tilde{a}$.  For such a deformation, we can always find a corresponding coordinates transformation  $\mathcal{F}:x^\mu\to \tilde{x}^\mu(x)=x^\mu + \eta^\mu(x)$ on the whole Euclidean space-time such that $\mathcal{F}(\tilde{a})= a$ (see Fig \ref{the deformation of subregion-setup}).

Via diffeomorphism equivalence we can then trade the shape-deformation for a metric deformation $g_{\mu\nu}\to \tilde{g}_{\mu\nu}$ together with other sources $j\to \tilde{j}$ and write the reduced density matrix elements as:
\be
\langle \alpha|\rho_{\tilde{a}}(g,\;j) |\beta\rangle =\langle \tilde{\alpha} | \rho_a(\tilde{g},\;\tilde{j}) |\tilde{\beta}\rangle  = \int_{\phi\left(a^{-}\right)=\tilde{\beta} }^{\phi\left(a^{+}\right)=\tilde{\alpha}}[\mathcal{D} \phi] e^{-I^{bulk}_{E}\left(\tilde{g},\;\tilde{j},\;\phi \right)}
\ee
where the new quantities are:
\be
\tilde{\alpha} = \alpha\circ \mathcal{F}^{-1},\;\;\tilde{\beta} = \beta\circ \mathcal{F}^{-1},\;\; \tilde{g}=g \circ \mathcal{F}^{-1},\;\;\tilde{j}=j\circ\mathcal{F}^{-1}
\ee
For simplicity we will assume $j=0$ and only consider bulk metric $g$.  For deformations $\eta^\mu(x)$ that are small, we can extract the change in the reduced density matrix to leading order in $\eta$. They are simply given by the linear response of the metric deformation, \ie proportional to the bulk stress tensor $T^{bulk}_{\mu\nu}$:
\bea
\delta \rho_a &=& U^\dagger\circ\rho_{\tilde{a}}\circ U -\rho_a = \frac{1}{2}\int \delta g^{\mu\nu} \rho_a\;\hat{T}^{bulk}_{\mu\nu} + \mathcal{O}\left(\zeta^2\right)\nonumber\\
\delta g^{\mu\nu} &=& 2\nabla^\mu\eta^\nu,\;\;U |\tilde{\alpha} \rangle = |\alpha\rangle
 \label{shape-deformation}
\eea
In order to compute the change in entanglement entropy, recall that under any change of the reduced density matrix: $\tilde{\rho}_a \to \rho_a + \lambda\; \delta \rho_a,\;\lambda \ll 1$, the first order change is dictated by the so-called entanglement first law:
\be
\delta S_{bulk}(a) = \lambda\; \text{tr}\delta \rho_a H_{a} + \mathcal{O}(\lambda^2),\;\;H_a = -\ln{\rho_a}
\ee
where $H_a$ is the (bulk) modular Hamiltonian associated with $\rho_a$. This is a consequence of the positivity constraint for the relative entropy between $\rho_a$ and $\tilde{\rho}_{a}$. Combining the entanglement first law with the shape perturbation theory \eqref{shape-deformation}, one can write the linear response of entanglement entropy under shape deformation in terms of correlation functions between stress tensor and modular Hamiltonian:
\bea
\delta S_{bulk}(a) &=& S_{bulk}(\tilde{a})-S_{bulk}(a) = -\text{tr}\left(U^\dagger \circ \rho_{\tilde{a}} \circ U\right) \ln{\left(U^\dagger \circ\ln{\rho_{\tilde{a}}}\circ U\right)} + \text{tr}\rho_a \ln{\rho_a}\nonumber\\
&=& \text{tr}\; \delta \rho_a\; H_a + \mathcal{O}\left(\eta^2\right) = \int_{\mathcal{B}} dx \sqrt{g} \,\  \nabla^\mu \eta^\nu(x)\; \text{tr} \left\lbrace \rho_a T^{bulk}_{\mu\nu}(x) H_a\right\rbrace + \mathcal{O}(\eta^2)\nonumber\\
&=& \int_{\mathcal{B}} dx \sqrt{g}\,\ \nabla^\mu \eta^\nu(x)\; \left\langle T^{bulk}_{\mu\nu}(x) H_a\right\rangle + \mathcal{O}(\eta^2)
\label{deformation-int}
\eea

\subsubsection{Extremization of bulk entropy by RT surface}
\begin{figure}[h]
     \centering
     \includegraphics[scale=0.45]{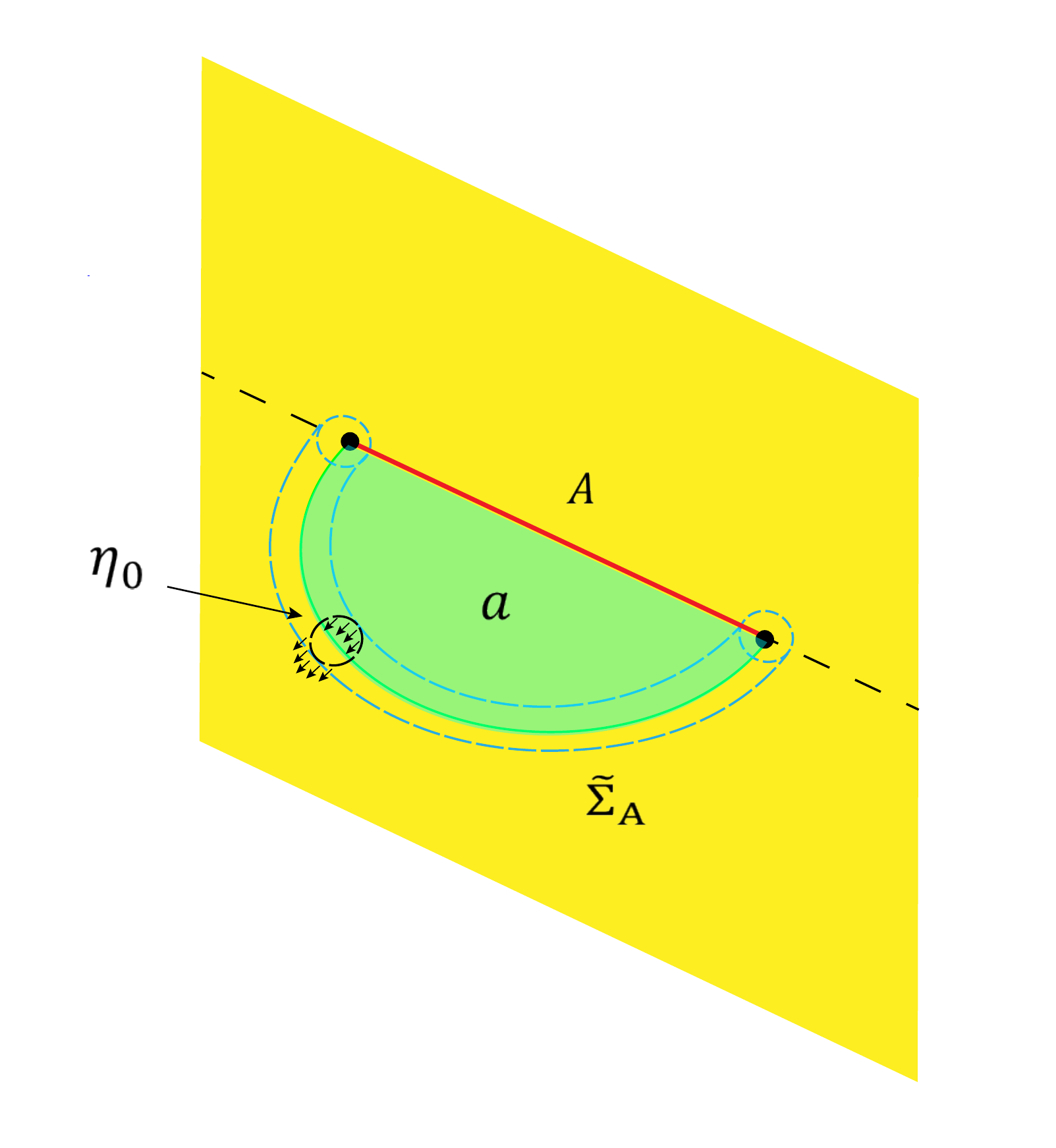}
     \caption{The UV cut-off tube $\partial \mathcal{B}= \mathds{S}^{\epsilon}_1 \times \tilde{\Sigma}_A$, outlined by the dotted blue lines. Only the zero-modes $\eta_0$ of the deformation vector fields are relevant for our analysis.}
     \label{fig:cut_off}
 \end{figure}

Now we use the entanglement first law (\ref{deformation-int}) to compute the shape response of bulk entanglement entropy across the RT surface $\tilde{\Sigma}_A$ in the cases with local modular flow. We will continue to work with the assumption that the local nature implies the existence of a $U(1)$ group of bulk isometries in the Euclidean bulk saddle $\mathcal{B}$, by which the RT surface is fixed under. Now we put in an arbitrary infinitesimal shape deformation about $\tilde{\Sigma}_A$ represented by the vector field $\eta^\mu(x)$ supported away from the asymptotic boundary. One can now integrate by parts of (\ref{deformation-int}) and write the linear response for the bulk entanglement entropy as a sum of boundary and bulk contributions:  
\be
\delta S_{bulk}(a) = \oint_{\partial \mathcal{B}} n^\mu \eta^\nu(x) \left\langle T^{bulk}_{\mu\nu}(x) H_a\right\rangle  + \int_{\mathcal{B}} dx \sqrt{g}\; \eta^\nu(x)\left\langle \nabla^\mu T^{bulk}_{\mu\nu}(x) H_a\right\rangle 
\ee
where $n^\mu$ is the unit normal vector to $\partial \mathcal{B}$. The second bulk term will vanish for general $x$ by the conservation law $\nabla^\mu T^{bulk}_{\mu\nu}(x)=0$  except for contact terms when $x$ coincides with the support of $H_a$, i.e. $x\in a$. When this happens, using the invariance of $H_a$ under and the local property of modular flow we can relate the correlation function at $x$ to that at any $x(\theta')\notin a$ and argue that they still vanish: 
\bea\label{eq:conservation}
&&\left\langle \nabla^\mu T^{bulk}_{\mu\nu}(x) H_a\right\rangle  = \text{tr} \rho_a \left( \nabla^\mu T^{bulk}_{\mu\nu}(x)H_a \right) \nonumber\\
&=&  \text{tr} \rho_a \left(e^{H_a \theta'}\; \nabla^\mu T^{bulk}_{\mu\nu}(x)\; H_a\; e^{-H_a \theta'}\right) 
=  \text{tr} \rho_a  \left(e^{H_a \theta'}\;  \nabla^\mu T^{bulk}_{\mu\nu}(x)\; e^{-H_a \theta'} \; H_a \right) \nonumber\\
&=&  \left[\frac{\partial x (\theta')}{\partial x}\right]^\alpha_\nu \text{tr} \rho_a  \left( \nabla^\mu T^{bulk}_{\mu\alpha}\left(x(\theta')\right) H_a\right) =0
\eea
So we are only left with the boundary term on $\partial \mathcal{B}$. The boundary consists of two components: one is the asymptotic boundary, but since we have made the support of $\eta^\mu(x)$ away from it so we do not need to consider this as part of $\partial \mathcal{B}$ for our purpose; what remains is a ``tube"  of small radius $\epsilon$ about the RT surface, see Fig (\ref{fig:cut_off}): 
\be
\partial \mathcal{B} = \mathds{S}^{\epsilon}_1 \times \tilde{\Sigma}_A 
\ee
The inclusion of this tube in $\partial \mathcal{B}$ is related to the UV regularization needed to compute entanglement entropy, in particular through the replica trick. The small radius $\epsilon$ is the UV cut-off parameter that we should send to $\epsilon \to 0$ in the end. 

In our cases, there is a $U(1)$ group of bulk isometries fixing $\tilde{\Sigma}(A)$. We invoke the Gaussian normal coordinates (\ref{eq:GN_E}) for the Euclidean metric near the fixed points $\tilde{\Sigma}(A)$ which is represented by $r=0$. In fact, the bulk isometries imply that the orthogonal radial geodesics never form caustics in the bulk, analogous to the null geodesics in the Lorentzian section; as a result the Gaussian normal coordinates (\ref{eq:GN_E}) is valid for the entire bulk. In order for the UV cut-off to reflect the underlying isometries, we take the cut-off tube to be located at constant $r=\epsilon$: 
\be
\partial \mathcal{B} = \left\lbrace \left(r=\epsilon,\theta, y\right), 0\leq \theta\leq 2\pi, y\in \tilde{\Sigma}_A \right\rbrace 
\ee
This way $\partial \mathcal{B}$ coincides with the image of modular trajectories. The vector field $\eta^\mu(x)$ is smooth at $r=0$, so using the complex coordinates $w=r e^{i\theta},\;\bar{w} = r e^{-i\theta}$,  we can decompose it into Fourier modes $\eta^\mu_n(r,y)$ each of which is smooth at $r=0$:
\be
\eta^{\mu}(x) = \sum_{n\geq 0} \eta^{\mu}_n(r,y)\left(w^n+\bar{w}^n\right),\;\;\eta^\mu_n(r,y)= \eta^\mu_n(0,y)+ r \left[\partial_r\eta^{\mu}_n\right](0,y) + ...
\ee
Different modes $\eta^{\mu}_n(r,y)$ deform properties of the tube at about $r=\epsilon$ in different ways. In particular, the zero-mode $\eta^\mu_0$ translate the tube locally at each point $y\in \tilde{\Sigma}_A$; while higher modes deform the shape of the tube locally, e.g. $\eta^\mu_1$ changes the radius of the tube, etc.  We are only interested in deforming the shape of $\tilde{\Sigma}_A$ but keeping the UV cut-off scheme intact, this corresponds to the zero-mode action $\eta^{\mu}_0$. Therefore for our purpose we shall work with the ansatz where all the higher modes are turned off: $ \eta^\mu_n=0, \; n\geq 1$, i.e. $\eta^\mu$ is independent of $\theta$. Furthermore, for the same shape deformation there is still gauge choice regarding its representation in terms of $\eta^\mu$, the most convenient gauge is so that it is orthogonal to the un-deformed surface $\tilde{\Sigma}_A$. In the coordinates (\ref{eq:GN_E}) this means it only has $w,\bar{w}$ components, and we can further set $\eta^w = \eta^{\bar{w}}=\eta$ so that the deformation respects reflection symmetry and remains on the $\tau=0$ plane, i.e. is purely spatial. Putting all these into consideration, $\delta S_{bulk}$ is computed by:
\bea
\delta S_{bulk}(a) &=& \int_{\tilde{\Sigma}_A} dy\;\eta (\epsilon,y)\sqrt{G_{ij}(\epsilon,y)}  G_{\theta\theta}(\epsilon,y)^{1/2}\epsilon^{-1} \Big(\oint_{|w|=\epsilon} dw\;\left\langle T^{bulk}_{ww}(w,\bar{w},y)\;H_a\right\rangle\nonumber\\
&+& \oint_{|w|=\epsilon} d\bar{w}\;\left\langle T^{bulk}_{\bar{w}w}(w,\bar{w},y)\;H_a\right\rangle\Big) + c.c
\eea
where being a zero-mode $\eta=\eta_0$ has been taken outside of the angular integrals. In the limit of sending UV cut-off $\epsilon \to 0$, the diverging $\epsilon^{-1}$ factor is cancelled by $G_{\theta\theta}(r,y)\to r^2$ behavior of metric as $r\to 0$. Now in order to argue that $\delta S_{bulk}=0$, we focus on one of the contour integrals $\oint w$ in the bracket and massage its form using its invariance under local modular flow as in (\ref{eq:conservation}):
\bea\label{eq:quantum_argument}
\mathcal{F}&=&\oint_{|w|=\epsilon} dw\; \left\langle T^{bulk}_{ww}(w,\bar{w},y)\;H_a\right\rangle = \oint_{|w|=\epsilon} dw\; \left\langle e^{H_a\theta'} T^{bulk}_{ww}(w,\bar{w},y) e^{-H_a \theta'}\;H_a\right\rangle\nonumber\\
&=&\oint_{|w|=\epsilon} dw\;e^{2i\theta'} \left\langle T^{bulk}_{ww}(w e^{i\theta'},\bar{w}e^{-i\theta'},y) \;H_a\right\rangle\nonumber\\
&=& e^{i\theta} \oint_{|w'|=\epsilon} dw'\; \left\langle T^{bulk}_{ww}(w',\bar{w}',y)\;H_a\right\rangle = e^{i\theta'}\mathcal{F}
\eea
where in the last line we have re-written the contour integral in terms of $w'=w e^{i\theta'}$. This is valid for any choice of $0\leq \theta'\leq 2\pi$, as a result we must have that $\mathcal{F}=0$. Using similar arguments one can easily show that the remaining contour integrals all vanish, and so $\delta S_{bulk}(a)=0$. It is worth pointing out that in this argument, the fact that it is an exact isometry instead of a conformal isometry is important, because otherwise there will be an additional $\theta'$-dependent conformal factor from  applying modular flow in the second line of (\ref{eq:quantum_argument}) that ruins the argument. In fact this is the origin of the non-zero response for the non-universal part of entanglement entropy across spheres in vacuum CFTs. 

We conclude that in the presence of $U(1)$ group of bulk isometries, the linear response of bulk entanglement entropy across the fixed-surface $\tilde{\Sigma}_A$ must vanish. As a result the bulk entanglement entropy contribution to $S_{gen}$ does not modify the position of Q.E.S relative the RT surface, so after including quantum corrections we still have the coincidence between Q.E.S and the causal horizon $\Sigma_A = \tilde{\Sigma}_A = \mathcal{C}_A$, i.e. the causal shadow is degenerate. 

\vspace{5mm}

\section{Linear shape-dependence of the bulk mutual information }\label{sec:bulk_MI}
\begin{figure}[h]
    \centering
    \includegraphics[scale=0.60]{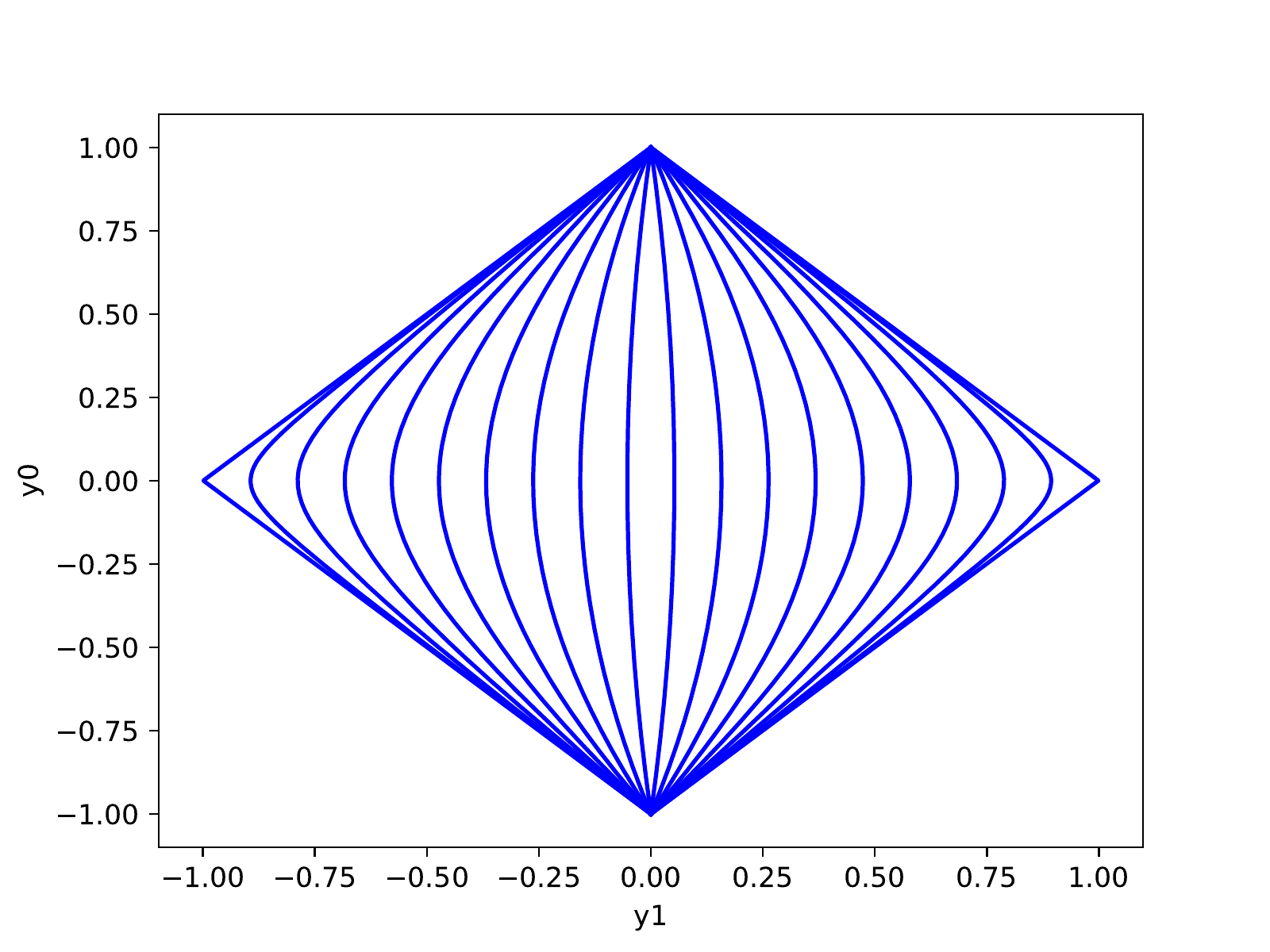}
    \caption{ An illustration for the trajectories of one interval modular flow. Where $y_{0}$ represents the time direction and $ y_{1} $ represents the space direction. As one can verify, these trajectories are just the conformal isometries of the corresponding CFT. }
    \label{one_interval_modular}
\end{figure}

From this section on we begin our analysis of causal shadows in the context of perturbation theory. The first goal is to work out the perturbation theory explicitly in a particular case. To this end we shall focus on the following set-up. The degenerate case we choose to perturb about is a spherical subregion $A$ of radius $R$ in the boundary CFT vacuum state. The conformal isometries are given by \cite{Casini2011May}, see Fig (\ref{one_interval_modular}): 
 \begin{equation}\label{eq:sphere_mod}
     y_s^{\pm}=R \frac{\left(R+y^{\pm}\right)-e^{\mp 2 \pi s}\left(R-y^{\pm}\right)}{\left(R+y^{\pm}\right)+e^{\mp 2 \pi s}\left(R-y^{\pm}\right)}
 \end{equation}
 \\
where we have taken the center to sit at the origin. There are various ways to perturb this case. In this paper, we take two spherical subregions $(A_1, A_2)$ of radii $(R_1, R_2)$ and place them in the same CFT vacuum separated by a large distance $L\gg R_1, R_2$. This can be viewed as a perturbation to two independent copies of the degenerate case: $\rho_{A_1\cup A_2} = \rho_{A_1} \otimes \rho_{A_2}$, which can be identified with the limit of $L=\infty$. At finite $L$, the density matrix does not factorize but receive corrections suppressed by the small conformal ratio $\lambda = 4 R_1 R_2/L^2$: 
\be
\rho_{A_1\cup A_2} = \rho_{A_1} \otimes \rho_{A_2} + \delta \rho_{A_1,A_2}(\lambda),\;\; \lim_{\lambda \to 0} \delta \rho_{A_1, A_2}(\lambda) \to 0 
\ee
From the modular flow point of view, $\rho_{A_1}$ and $\rho_{A_2}$ generate independent local modular flows, represented by conformal isometry groups fixing $\partial A_1$ and $\partial A_2$ respectively. When we couple the two subregions by placing them in the same CFT vacuum and separated by a finite $L$, the conformal isometries become incompatible with one another: the integral curves, or modular trajectories will cross. Therefore for finite $L$ the conformal isometries are broken and we no longer have smooth local modular flows about $A_1 \cup A_2$. 

As a consequence, in the bulk we can no longer argue the coincidence between the Q.E.S and causal horizon. At the classical level, since we are far away from the entanglement transition, the RT surface $\tilde{\Sigma}_{A_1\cup A_2}$ is simply the union of two RT surfaces associated with $A_1$ and $A_2$ individually: 
\be
\tilde{\Sigma}_{A_1\cup A_2}= \tilde{\Sigma}_{A_1} \cup \tilde{\Sigma}_{A_2} 
\ee
For disconnected $A_1$ and $A_2$ their causal horizons do not affect one another:
\be
\mathcal{C}_{A_1\cup A_2} = \mathcal{C}_{A_1} \cup \mathcal{C}_{A_2}
\ee
Therefore from the discussion of section (\ref{subsec:classic}) we have that $\tilde{\Sigma}_{A_1\cup A_2} = \mathcal{C}_{A_1\cup A_2}$, and the causal shadow is still degenerate classically. The finite $\lambda$ effects only show up at the level of bulk quantum corrections. In particular, the bulk generalized entropy is contributed by the bulk entanglement entropy across both $\tilde{\Sigma}_{A_1}$ and $\tilde{\Sigma}_{A_2}$ which is not simply additive: 
\be
S_{bulk}(a_1 \cup a_2) \neq S_{bulk}(a_1) + S_{bulk}(a_2) 
\ee
As a result, the extremal property of $S_{bulk}$ under shape deformation is broken by including both $a_1$ and $a_2$ as the subregion in the bulk. Therefore we expect non-zero corrections to the location of the Q.E.S relative to the RT surfaces while the causal horizon is defined purely classically and thus remains intact. In this way a causal shadow is generated perturbatively from the quantum corrections. Indeed from the discussion of section (\ref{subsec:quantum}), the non-extremality under shape deformation is entirely contained in the bulk mutual information term: 
\be\label{eq:MI_nv}
\delta S_{bulk}(a_1\cup a_2) =   \delta S_{bulk}(a_1) + \delta S_{bulk}(a_2) - \delta S_{bulk}(a_1\cup a_2) = \delta I_{bulk}(a_1, a_2) \neq 0 
\ee
To declutter notations we omit the bulk subscript from now on, all quantities are of bulk nature unless specified otherwise. Therefore in order to capture the causal shadow in this case, the first step is to grasp the non-vanishing response of $\delta I$ under shape deformations. It acts like the source that supports the genesis of the causal shadow. In the remaining of this section, we apply the shape perturbation theory and entanglement first law as in (\ref{subsec:quantum}) to compute the linear shape dependence of bulk mutual information, this time without the bulk isometries. 

\subsection{Operator product expansion of the bulk modular Hamiltonian}
The crucial element for applying the entanglement first law is the knowledge about the un-deformed modular Hamiltonian. In the present case this is the bulk modular Hamiltonian associated with the two bulk subregions $H_{a_1\cup a_2}$. The corresponding modular Hamiltonian from each component is fixed by the bulk isometries to be: 
\be
H_{a_i} = \int_{a_i} dx q^\mu_i(x)n^\nu(x)T_{\mu\nu}(x),\;\;i=1,2
\ee
where $q^\mu_i$ is the generator of the bulk isometry associated with $a_i$ and $n^\mu$ is the bulk normal vector to the time-slice. They do not produce a linear response under shape deformations, so as suggested by (\ref{eq:MI_nv}) we can subtract them off $H_{a_1\cup a_2}$ and only consider the remaining part associated with the mutual information $I(a_1,a_2)$: 
\be 
\Delta H_{a_1\cup a_2} = H_{a_1\cup a_2} - H_{a_1} -H_{a_2}
\ee 
Apart from a few special cases e.g. the free Dirac fermion in two dimensions \cite{Casini2005Jul, Casini2009Mar,Casini2009Sep}, the modular Hamiltonian for the subregion consisting of multiple spheres is hard to obtain in general. Using the replica trick, the mutual information can be obtained from the correlation function of two twist operators in the orbifold QFTs: 
\be 
I(a_1\cup a_2) = \lim_{n\to 1}\frac{1}{1-n}\ln{\left[\frac{\langle \Sigma^{a_1}_n \Sigma^{a_2}_n\rangle}{\langle \Sigma^{a_1}_n\rangle \langle \Sigma^{a_2}_n\rangle}\right]}
\ee
where $\Sigma^{a_{1,2}}_n$ are the twist operators inserted across $\partial_{a_{1,2}}$. These are extended operators in the orbifold QFTs for spacetime dimensions $d>2$. In the large distance limit, we can resort to the OPE approach and extract the leading order results. This means that we can expand the twist operators in terms of the local operator content in the orbifold QFTs. The OPE results of the mutual information between two spheres in the CFT vacuum has been studied in \cite{Cardy2013,Faulkner2016Aug, ChenBin2017} which take the form: 
\be 
I_{a_1\cup a_2} \sim \left(\frac{4R_1 R_2}{L^2}\right)^{2\Delta} +...\sim \lambda^{2\Delta} +...
\ee
where $\Delta$ is the lowest conformal dimension in the operator spectrum of the CFT. The more refined OPE at the level of the modular Hamiltonian for $\Delta H$ has been similarly worked out in \cite{Faulkner2021Aug} in general terms, which we can adopt and apply to the computation of $\Delta H_{a_1\cup a_2}$. 

The bulk modular Hamiltonian $H_{a_1\cup a_2}$ is defined with respect to the bulk effective QFT, which is not a CFT. Despite this the OPE method is still useful if we are only interested in obtaining the leading order result in the large separation limit $\lambda\ll 1$, and also particularly at the leading order in $1/N$, where the bulk effective QFT consists of free fields with various masses and spins. At this order the bulk OPE computation can be performed explicitly, as was done in \cite{Faulkner2016Aug}, and one can check that the OPE takes the proper form where higher order contributions are suppressed by higher powers of $\lambda$. The bulk OPE for $\Delta H_{a_1\cup a_2}$ proceeds by summing over contributions from these free fields. The dominant OPE contribution comes from that of the smallest mass, which for simplicity we assume to be a scalar. To be more concrete, we consider the relevant bulk action to consists of a massive free scalar minimally coupled to the bulk metric:
   \begin{equation}
       \begin{aligned}
           S &= S_{g} + S_{matter} \\
           &=  \frac{1}{16 \pi G_{N}} \int_{\mathcal{B}} d^{d+1}{x} \sqrt{g} \left( R -2 \Lambda \right) +   \int_{\mathcal{B}} d^{d+1}{x} \sqrt{g} \left( -\frac{1}{2}g^{\mu \nu} \nabla_{\mu}\phi \nabla_{\nu}\phi - \frac{1}{2} m^{2} \phi^{2} \right)
       \end{aligned}
      \label{the-action}
   \end{equation}
   where $\Lambda = - \frac{d(d-1)}{2}$ is the cosmological constant and $\phi$ is scalar field with mass $m$. Only the matter action $S_{matter}$ is relevant for $\Delta H_{a_1\cup a_2}$, the bulk metric can be taken as a fixed $\text{AdS}_{d+1}$ background, which we write in Poincar{\'e} coordinates:
  	\begin{equation}
      ds^{2} = \frac{-dt^2+\sum_{i=1}^{d-1}dx_{i}^{2} +dz^{2} }{z^{2}},\quad z>0
  \end{equation}
In this coordinate, the bulk subregions $a_{1}$ and $a_{2} $ are spatial hemispheres on the bulk $t=0$ slice with radii $R_{1}$ and $R_{2}$ about the boundary centers $(r_1,\epsilon_z)$ and $(r_2,\epsilon_z)$, where $\epsilon_z \to 0$ is the UV regulator for their position along the $z$ direction: 
\be
a_1 = \lbrace z^2+(x-r_1)^2 =R_1^2,\;\; z>0\rbrace,\;\; a_2 = \lbrace z^2+(x-r_2)^2 =R_2^2,\;\; z>0\rbrace 
\ee
We need to perform the OPE for the bulk twist operators $\Sigma^{a_{1,2}}_n$ into local operators of the $\phi$ sector in the orbifold bulk effective QFT, i.e. $\phi^{(i)}, i=1,...,n$. A convenient choice for the location of $\phi^{(i)}$ when expanding $\Sigma^{a_{1,2}}_n$ is about their centers $(r_{1,2},\epsilon_z)$. It turns out that the leading order OPE takes the form of the following bi-local operators connecting the different replica: 
 \begin{equation}\label{eq:twist_OPE}
     \Sigma_{n}^{a_{1}, a_{2}}=\left\langle \Sigma_{n}^{a_{1}, a_{2}} \right\rangle_{\mathcal{B}} \left(1+\frac{1}{2}\left(2 R_{1, 2}\right)^{2 \Delta} \sum_{j \neq k}^{n-1} c^n_{j-k} \phi^{(j)}(r_{1,2}) \phi^{(k)}(r_{1,2},\epsilon_z)+\cdots\right)
 \end{equation}
where the scaling dimension $\Delta$ is determined by $m^{2} =\Delta (\Delta-d) $. Similar to the usual OPE, the coefficients $c_{j-k}$ are obtained by taking the limit: 
\be
c^n_{j-k} =\lim_{r\to 0} \frac{\big\langle \Sigma^{a_{1,2}}_n \phi^{j}(r)\phi^{k}(r)\big\rangle}{\big\langle \phi(r_{1,2},\epsilon_z)\phi(r)\big\rangle^2 }
\ee
Using the bulk isometries across the individual hemi-sphere $a_{1,2}$ in the Euclidean $\text{AdS}_{d+1}$ background, $c_{j-k}$ can be explicitly obtained in \cite{Faulkner2016Aug} from the thermal correlator $g_n(\tau)$ between $\phi(0)$ and $\phi(\tau)$ on a bulk hyperbolic manifold with a conical defect:
 \begin{equation}
     \begin{aligned}
         c^n_{j-k} 
         &= \epsilon_z^{-2\Delta} \left(\frac{2\Delta -d}{C_\Delta}\right) g_{n}\left(\tau_{j}-\tau_{k}\right), \; \tau_{i}=(2 i+1) \pi,\;\;C_\Delta = \frac{\Gamma(\Delta)}{\pi^{d/2}\Gamma(\Delta -d/2)}
     \end{aligned}
 \end{equation} 
One can make use of the twist operator OPE (\ref{eq:twist_OPE}) to derive that of the modular Hamiltonian term $\Delta H_{a_1\cup a_2}$. This has been done in \cite{Faulkner2021Aug}, and we briefly describe the derivation below. The main idea is still to use the replica trick, but this time start by computing the matrix elements of $\rho^n_{a_1\cup a_2} \propto \Sigma^{a_1}_n \Sigma^{a_2}_n$. The first step is to perform OPE on one of the twist operators, say $\Sigma^{a_2}_n$: 
\be\label{eq:renyi_OPE}
\rho^n_{a_1\cup a_2} \propto \Sigma^{a_1}_n \left(1+\frac{1}{2}(2R_2)^{2\Delta} \sum^{n-1}_{j\neq k} c^n_{j-k}\phi^{(j)}(r_2,\epsilon_z)\phi^{(k)}(r_2,\epsilon_z)+...\right)
\ee
which converges well in the limit of $\lambda \ll 1$. The next step is to compute the matrix elements $\langle \psi_1|\rho^n_{a_1\cup a_2}|\psi_2\rangle$ via the OPE (\ref{eq:renyi_OPE}). This can be realized as a Euclidean path-integral on the conifold across $a_1$ with operators $\phi^{(j)}\phi^{(k)}$ as well as $\psi_{1,2}$ inserted, the latter two located only on one of the branches. The summation can be implemented by contour integrals in the modular flow parameters $s_j=2\pi i j,\;s_k = 2\pi i k$. After appropriately deforming the integration contours, they can be brought into forms ready for the analytic continuation $n\to 1$. In this process the modular flow parameters $s_j,s_k$ are lifted onto the real sections. We refer interested readers to \cite{Faulkner2021Aug} for more details. At the end of the day, we obtain the OPE for $\Delta H_{a_1\cup a_2}$ to be of the form:
    \begin{equation}\label{eq:MI_mod}
        \begin{aligned}
            \Delta H_{a_{1} \cup a_{2}} &=  \epsilon_z^{-2\Delta} \bigg( \frac{2 \Delta-d}{ C_{\Delta}} \bigg)  \Bigg( -\lambda^{2 \Delta}                \int_{-\infty}^{\infty} {d s}\,\  k_{1}(s) \;   \phi^{-is}\left(r_2, \epsilon_z \right) \phi\left(r_{2}, \epsilon_z\right)+\frac{i \lambda^{2 \Delta}}{2 \pi} \\
            & \times \int_{-\infty}^{\infty} {d s_{j} d s_{k} } \,\  k_{2}(s_{j}, s_{k}) \; \phi^{-is_k-is_j}\left(r_2, \epsilon_z \right) \phi^{-is_j}\left(r_2, \epsilon_z\right) \Bigg) \\
        \end{aligned}
    \end{equation}
where the integration kernels $k_1(s)$ and $k_2(s_j,s_k)$ are specified by:
 \begin{equation}
     \begin{aligned}
         k_{1}(s) &= \frac{g_{1}(-is + \pi)}{4\cosh^{2}(s/2)}  \\
         k_{2}(s_{j}, s_{k}) &= \frac{1}{ 4\cosh^{2}(s_{j}/2)}\bigg( \frac{1}{e^{s_{k} + i\epsilon}-1 } + \frac{1}{e^{s_{k} + s_{j}}+ 1 }\bigg) c_{1}(- i s_{k} + \epsilon) \\
         c_{1}(\theta) &= \frac{1}{(2-2 \cos \theta)^{\Delta}} 
     \end{aligned}
 \end{equation}
We mention that $ g_{1}( \theta)$ is just the analytical continuation of the OPE coefficient $g_n$ at $n=1$. The operator $\phi^{-is}(r_2, \epsilon_z)$ is defined by the bulk modular flow using $\rho_{a_1}$:
 \begin{equation}
     \phi^{-is}(r_2, \epsilon_z) = \rho_{a_1}^{is -1/2} \phi(r_{2}, \epsilon_z) \rho_{a_1}^{- is + 1/2} 
 \end{equation}
Notice that the additional $(\pm 1/2)$ term in the exponent of $\rho_{a_1}$ means that $\phi^{-is}(r_2,z)$ is in fact an element of the operator algebra in $D(a_1)$. By combining the $\epsilon_z ^{-2\Delta}$ factor with the bulk operators $\phi(r_2, \epsilon_z)$ in (\ref{eq:MI_mod}), one recovers the boundary operator $\mathcal{O}$ dual to $\phi$ via the extrapolate dictionary: 
\be
\lim_{\epsilon_{z}\to 0}\epsilon_z^{-\Delta} \phi(x,\epsilon_z) \to \mathcal{O}(x) 
\ee
Then by further identifying the bulk modular flow by $\rho_{a_1}$ with the boundary modular flow by $\rho_{A_1}$ through the JLMS proposal, one realizes that $\Delta H_{a_1\cup a_2}$ in fact matches to its boundary counterpart, see \cite{Chen2022}: 
\be
\Delta H^{bulk}_{a_1 \cup a_2} = \Delta H^{bdry}_{A_1\cup A_2} 
\ee 
 
\subsection{Linear response of the bulk mutual information under shape deformations} 
With the OPE result (\ref{eq:MI_mod}) for the modular Hamiltonian $\Delta H_{a_1\cup a_2}$, we can go straight ahead and apply the entanglement first law to compute the linear response of the bulk mutual information under the shape deformation. Since we have chosen the OPE for $\Sigma^{a_2}_n$ when expanding $\Delta H_{a_1\cup a_2}$, it makes sense to focus only on deforming $a_1$. They are defined similarly as before by a deformation vector field $\eta$ on the RT surface: 
\be
\Sigma(A_1) \to \Sigma(A_1) + \eta 
\ee
The application of the entanglement first law and shape perturbation theory proceeds the same as in the subsection (\ref{subsec:quantum}), in the end we obtain the variation of the bulk mutual information under $\eta$ as the following surface integral: 
  \begin{equation}\label{eq:MI_1st}
      \begin{aligned}
          \delta I(a_{1}, a_{2}) 
          &=  \int_{\partial \mathcal{B}} {d}^{d} Y \sqrt{h(Y)} \,\ n^{\mu} \eta^{\nu}(Y)  \langle  T^\phi_{\mu \nu}(Y) \Delta H_{a_{1} \cup a_{2} } \rangle
      \end{aligned}
  \end{equation}
where $T^\phi_{\mu\nu}$ is the matter part stress tensor of the bulk action \eqref{the-action}: 
\be\label{eq:bulk_T}
T^\phi_{\mu\nu} = \nabla_{\mu}\phi \nabla_{\nu}\phi - \frac{1}{2}g_{\mu \nu} \left(\nabla\phi\right)^{2}-\frac{1}{2}g_{\mu \nu} m^{2} \phi^{2}
\ee
and $\partial \mathcal{B}$ is basically the same cut-off tube discussed before, see Fig (\ref{fig:cut_off}). One then needs to plug (\ref{eq:MI_mod}) and (\ref{eq:bulk_T}) into (\ref{eq:MI_1st}) and crank through the subsequent computations. This is tedious but relatively straightforward. An analogous computation was performed in \cite{Chen2022} for CFTs. The basic ingredients are correlators of the form: 
\be\label{eq:Tff}
\mathcal{F}_{\mu\nu}(Y,s,s')=\epsilon_z^{-2\Delta}\langle T^\phi_{\mu\nu}(Y) \phi^{-is}(r_2,\epsilon_z) \phi^{-is'}(r_2,\epsilon_z)\rangle 
\ee
integrated over $(x,s,s')$ with the appropriate kernels. While the CFT counterpart $\langle T_{\mu\nu}\mathcal{O}\mathcal{O}\rangle$ in \cite{Chen2022} is fixed by conformal ward identity, (\ref{eq:Tff}) is evaluated at the leading order in $1/N$ via wick-contraction of the bulk free field $\phi$: 
\be
\mathcal{F}_{\mu\nu}(Y,s,s') \propto \hat{\mathcal{D}}_{\mu\nu}G^\Delta_{b\partial}(Y;r^s_2) G^\Delta_{b\partial}(Y;r^{s'}_2)
\ee
where $\hat{\mathcal{D}}_{\mu\nu}$ is the appropriate differential operator with respect to $Y$ defined by (\ref{eq:bulk_T}), and $G^\Delta_{b\partial}$ is the bulk-boundary propagator in $\text{AdS}_{d+1}$:
\be 
G^\Delta_{b\partial}(Y;r^s_2)=  \frac{ C_{\Delta}}{2 \Delta-d} \left[\frac{Z}{Z^2+(Y_{\partial}-r^{s}_2)^2}\right]^{\Delta}
\ee
written in Poincar{\'e} coordinates denoted by $Y=(Y^0,Y^1,...,Y^d)=(T,Y^{i\geq 1},Z) =(Y_{\partial},Z)$. The remaining boundary coordinate $r^s_2$ is defined by the local modular flow using $\rho_{A_1}$:
\be\label{eq:flow_def}
\rho_{A_1}^{-is+1/2}\mathcal{O}(r_2)\rho^{is-1/2}_{A_1} \propto \mathcal{O}(r^s_2) 
\ee
The explicit expression for $r^s_2$ is given by the messy form (\ref{eq:sphere_mod}) for spherical $A_1$. To facilitate the computation we can use the bulk isometry: 
\bea \label{eq:map}
&&  Y^{\alpha} = \frac{\Lambda x^{\alpha} - C^{\alpha} \Lambda^{2} x^{2} }{ 1 - 2 \Lambda C\cdot x + \Lambda^{2} C^{2} x^{2}} +   2R_{1}^{2} C^{\alpha} \nonumber\\
&& C^{\alpha} = \frac{\delta_{1}^{\alpha}}{ 2 R_{1}} ,\quad  \Lambda = \frac{2R_{1} (L^{2} - R_{1}^{2} -R_{2}^{2})}{(L+R_{1} + R_{2}) (L+R_{1} - R_{2})}
\eea
to map between the hemi-sphere in $Y$ coordinates $\lbrace |Y|=R_1, T=0, Z>0\rbrace$ and the bulk Rindler half-space $\lbrace x^0=0, x^1<0, z>0\rbrace$ in $x=(x^i,z)$ coordinates. Under this mapping, the other hemi-sphere $a_2$ which in $Y$-coordinate is of radius $R_2$ centered at $(0,L,0,...,0)$ becomes a hemi-sphere of radius $\frac{\lambda}{2}  $ centered at $r_2=(0,1,0,...,0)$, where $\lambda = \frac{4 R_{1} R_{2}}{L^{2}}$ is the conformal ratio defined before. The modular flows in the $x$ coordinates then simply boost along the $(x^1,x^0)$ plane, so we define as before the holomorphic/anti-holomorphic coordinates $\lbrace w = x^{1} + ix^{0} = re^{i\theta},\bar{w}  = x^{1} - ix^{0} = re^{-i\theta}\rbrace$ which are moved under the modular flow (\ref{eq:flow_def}) according to:
\be
w^{is} = e^{s}w;\;\;\bar{w}^{is}=e^{-s}\bar{w};\;\;x^{is}_i =x_i,\;i\geq 2; \;\;z^{is}=z
\ee
From now on we will do everything in the Rindler coordinates $x$, including computing the shape of the causal shadow, and only map back to the spherical coordinates $Y$ in the end. Similar to the discussion in subsection (\ref{subsec:quantum}), regarding the shape-deformation vector field $\eta^\mu$ we only consider the zero-mode with components orthogonal to the Rindler plane: $\eta^0(w,\bar{w},z,x^i)=\eta(z,x^i)\left(\partial_w+\partial_{\bar{w}}\right)$. Putting things together, to obtain the response $\delta I_{a_1\cup a_2}$ by a local deformation $\delta \eta(z,x^i)$ on $\Sigma(A_1)$ we need to compute the following integral:   
\bea\label{var-MI-1}
&&\frac{\delta I(a_{1}, a_{2})}{\delta \eta(z,x^i)}= z^{-d}\; \oint_{|w|=\epsilon}  \,\ dn^{\mu} \langle T_{\mu w}(z, x_{i},w,\bar{w}) \Delta H_{a_{1} \cup a_{2} } \rangle + (T_{\mu w}\to T_{\mu\bar{w}}) \nonumber\\
 &=& z^{-d}\; \oint_{|w|=\epsilon}  \,\ dn^{\mu} \hat{\mathcal{D}}_{\mu w} \Big(\int^\infty_{-\infty} ds\; k_1(s) G_{b\partial}(z, x_{i},e^{s}w,e^{-s}\bar{w};-r_2) G_{b\partial}(z, x_{i},w,\bar{w};r_2) \nonumber\\
&+& \int^{\infty}_{-\infty}ds_j ds_k\;k_2(s_j,s_k) G_{b\partial}(z, x_{i},e^{s_j+s_k}w,e^{-s_j-s_k}\bar{w};-r_2) G_{b\partial}(z, x_{i},e^{s_j}w,e^{-s_j}\bar{w};-r_2)\Big)\nonumber\\
&+& (\hat{\mathcal{D}}_{\mu w}\to \hat{\mathcal{D}}_{\mu\bar{w}})
\eea
We refrain from going through the remaining steps of computing (\ref{var-MI-1}), they closely follow \cite{Chen2022}  which can be referred to for more details. Instead we make a few remarks below regarding the general patterns that emerge during the computation. 
\begin{itemize}
\item Each term in (\ref{var-MI-1}) contains one contour integrals $\oint dw$ or $\oint d\bar{w}$, which amounts to extracting the residue of simple poles for $w$ or $\bar{w}$ that emerge from integrating over the remaining modular parameters. 
\item Among the different terms in (\ref{var-MI-1}) labelled by the components of $\hat{\mathcal{D}}_{\mu\nu}$, only the $ww$ and $\bar{w}\bar{w}$ component terms contribute, and furthermore the contributions only come from the $\partial_\mu\partial_\nu$ part of $\hat{\mathcal{D}}_{\mu\nu}$. 
\item In terms of the remaining modular integrals, the modular Hamiltonian can be decomposed into two terms:
\be
\Delta H_{a_1\cup a_2} = \int^\infty_{-\infty} ds ... + \int^\infty_{-\infty}ds_j \int^\infty_{-\infty} ds_k ...
\ee
depending on whether it involves a single or double integral of the modular parameters; it turns out only the double integral $\int^\infty_{-\infty} ds_j \int^\infty_{-\infty} ds_k$ yields simple poles in $w$ or $\bar{w}$. 
\item The two integrals $\int^\infty_{-\infty} ds_j \int^\infty_{-\infty} ds_k$ play different roles, the $s_j$ integral gives rise to simple poles in $w$ or $\bar{w}$, whose $s_k$-dependent residues are then integrated over to give the final results for $\delta I/\delta \eta$. This means for example that one can change the order of integration between $\oint dw$ and $\int^\infty_{-\infty}ds_k$, i.e. the shape response of mutual information receives contributions additively from the $ds_k$ integral. We remind that $s_k$ measures the modular separation between the bi-local operator $\phi(-is_j-s_k)\phi(-is_j)$ in the OPE. 
\end{itemize}
At this point we simply write down the final result of computing (\ref{var-MI-1}) as follow:
\bea \label{var-MI-final}
 \frac{\delta I(a_{1}, a_{2})}{\delta \eta(z,x^i)} &=&    N_{d, \Delta}  z^{2\Delta-d}   (1+z^{2}+\sum_{i \geq 2} x_{i}^{2})^{-2\Delta-1} \nonumber\\
       N_{d, \Delta} &=&    \lambda^{2 \Delta }  \frac{ \pi^{3/2} }{4^{2\Delta} \pi^{d/2} } \frac{\Gamma( 2\Delta  +1 )^{2}}{\Gamma(\Delta )  \Gamma(2\Delta + 3/2)} \frac{1}{\Gamma(\Delta -d/2 + 1)}
\eea

\vspace{5mm}

\section{Causal shadows from quantum corrections to the Q.E.S}\label{sec:main}
In this section, we can use the final result (\ref{var-MI-final}) of the previous section to compute the shape of the causal shadow it induces as the quantum correction to the location of the Q.E.S. More precisely,  based on (\ref{var-MI-final}) we compute the contribution to the shape of the causal shadow from the OPE channel of a bulk scalar of mass $m^2 = \Delta (\Delta -d)$. We remind that the boundary subregion consists of two spheres $A_1$ and $A_2$ separated by large distance $L$ (See Fig \ref{two distant spheres}), and we focus on computing the Q.E.S component homologous to $A_1$ because we rely on the OPE of the other component $A_2$. The Q.E.S component homologous to $A_2$ can be computed via the other way of doing the OPE, and will be identical in shape to that of $A_1$ if $R_1=R_2$.  

As was discussed in section (\ref{sec:degenerate}), the classical RT surface $\tilde{\Sigma}(A_{1,2})$ solves the Q.E.S equation in the absence of the other component: 
\be\label{eq:QES_0th}
\frac{\delta S_{gen}(\Sigma)}{\delta \eta}\bigg|_{\tilde{\Sigma}(A_{1,2})}=\frac{\delta \text{Area}(\Sigma)}{4 G_N \delta \eta}\bigg|_{\tilde{\Sigma}(A_{1,2})}+ \frac{\delta S(a)}{\delta \eta}\bigg|_{\tilde{\Sigma}(A_{1,2})} =0
\ee 
The two terms on the RHS vanish separately. Now with both spheres present but separated by a large distance $L$, the Q.E.S receive quantum corrections from the correlation of these two spheres. This is encoded in the bulk mutual information $I$. The goal is then to solve the Q.E.S equation in the presence of the bulk mutual information. We can work with the perturbation theory and write the Q.E.S schematically as: 
\be\label{eq:QES}
\Sigma_{QES} = \Sigma_{RT} +\zeta,\;\;\Sigma_{RT} = \Sigma(A_1)\cup \Sigma(A_2)
\ee 
Analogous to $\eta$ when defining shape-responses before, $\zeta$ can be represented more precisely by a normal displacement field, with the sign chosen that $\zeta>0$ corresponds to deforming outward (away from the asymptotic boundary). We focus on the part of $\zeta$ supported on $\Sigma(A_1)$. One then proceeds by plugging (\ref{eq:QES}) into the extremality equation for the Q.E.S:
\be
\frac{\delta S_{gen}\left(\Sigma_{RT}+\zeta\right)}{\delta \eta(x)}=0,\;\;\forall\;x\in\Sigma_{RT}+\zeta
\ee 
This can then be expanded in $\zeta$. Taking into account of (\ref{eq:QES_0th}) one obtains the equation for $\zeta$ up to the linear order as follows: 
\be\label{eq:QES_1st}
\frac{ \delta I_{a_1\cup a_2} }{ \delta \eta(x)} \bigg|_{\Sigma^{\prime} = \Sigma_{RT}}+\frac{1}{4 G_N}   \int _{\Sigma_{RT}}{dy} \; \frac{\delta^{2} \operatorname{Area}\left(\Sigma^{\prime} \right)}{\delta \eta(y) \; \delta \eta(x) } \bigg|_{\Sigma^{\prime} = \Sigma_{RT}} \zeta(y) + \mathcal{O}(\zeta^2) =0,\;\;\forall x\in \Sigma_{RT}
\ee
This equation can in principle be inverted to solve $\zeta$ order by order in $G_N$, the leading order solution is determined by the linear truncation of (\ref{eq:QES_1st}):
    \begin{equation}\label{eq:CS_1st}
         \zeta(x)=4 G_N \int_{\Sigma_{RT}} d y \;  \mathcal{K}\left(x, y\right) \left[\frac{\delta I_{a_1\cup a_2}}{\delta \eta\left(y\right)}\right] + \mathcal{O}\left( G_{N}^{2} \right)
    \end{equation}
The factor $\left(\delta I_{a_1\cup a_2}/\delta \eta\right)$ is precisely the linear shape dependence of the mutual information that we have computed in section (\ref{sec:bulk_MI}), which is also expanded in $\lambda$ according to the OPE of $\Delta H_{a_1\cup a_2}$. So in terms of the actual computations (\ref{eq:CS_1st}) is a double expansion in both $G_N$ and $\lambda$. We are aiming to capture the leading order results in both expansions. The kernel $\mathcal{K}\left(x, y\right)$ is the inverse of the second order functional derivative of the area functional evaluated at $\Sigma_{RT}$ (where the first order derivatives vanish):
    \begin{equation}\label{eq:area_kernel}
       \mathcal{K}\left(x, y\right)= \left( \frac{\delta^2 \operatorname{Area}(\Sigma^{\prime})}{\delta \eta\left(x\right) \delta \eta(y)} \bigg|_{\Sigma^{\prime} = \Sigma_{RT}} \right)^{-1} 
    \end{equation}
Its matrix elements are labelled by the coordinates on $\Sigma_{RT}=\Sigma(A_1) \cup \Sigma(A_2)$, the geometric nature of (\ref{eq:area_kernel}) means that it has an obvious block diagonal structure: 
\be
\mathcal{K}(x,y) = 0,\;\;x\in \Sigma(A_1),\; y\in \Sigma(A_2) 
\ee   
Therefore for our purpose of computing $\zeta(x),\;x\in \Sigma(A_1)$ we can restrict the integral in (\ref{eq:QES_1st}) to only $y\in \Sigma(A_1)$. Via the bulk isometry (\ref{eq:map}), the classical RT surface $\Sigma(A_1)$ is the bulk Rindler half-plane $\lbrace x^1<0\rbrace$. We continue to work in the Rindler coordinates $x$, this which (\ref{eq:QES_1st}) takes the more explicit form:
\begin{equation}
           \zeta\left(z_{1}, x_{i}\right)=4 G_{N}  \int_{0}^{+\infty} {d z_{2}}  \int {d y_{j \ge 2}}  \;\;  \mathcal{K}\left(z_{1}, x_{i} ; z_{2}, y_{j} \right) \frac{\delta I_{a_1\cup a_2}}{\delta \eta\left(z_{2}, y_{j}\right)}+\mathcal{O}\left(G_{N}^{2}\right)
       \label{CS_profile}
\end{equation}
    $$
    \mathcal{K}\left(z_{1}, x_{i} ; z_{2}, y_{j} \right) =\left(\left.\frac{\delta^{2} \operatorname{Area}(\Sigma^{\prime})}{\delta \eta\left(z_{1}, x_{i}\right) \delta \eta\left(z_{2}, y_{j} \right)}\right|_{\Sigma^{\prime}=\Sigma_{R T}}\right)^{-1}
    $$
\vspace{1mm}
\subsection{Computing the kernels in Rindler coordinates}
In order to compute the shape of the perturbative causal shadow $\zeta(z,x_i)$ via (\ref{CS_profile}) we need to supply the explicit form of the kernel $  \mathcal{K}\left(z_{1}, x_{i} ; z_{2}, y_{j} \right) $. To this end, we need to solve the Green's function for the local fluctuating field $\eta(x)$, whose effective action is the area functional.  In more concrete terms, the ``vacuum" of the effective action is the extremal surface $\Sigma(A_1) : \lbrace x_0 = x_1=0, z,x_{i\geq 2}\rbrace$; the field $\eta$ parametrizes the fluctuating surface $\Sigma^{\prime}$ about $\Sigma(A_1)$ as $\Sigma^{\prime}:\lbrace x_0=0, x_{1}  = \eta(z, x_{i \ge 2}), z,x_{i\geq 2})$. We need to compute the area functional up to quadratic order in $\eta$. In Rindler coordinates this is given by:
    \begin{equation}
        \begin{aligned}
            S(\eta) =\operatorname{Area}(\Sigma^{\prime})  &= const +  \frac{1}{2}\int_{0}^{+\infty} \frac{d z}{z^{d-1}}  \int {d x_{ i \ge 2}} \; \bigg[  \Big( \frac{\partial \eta }{\partial z }  \Big)^{2} + \sum_{i \ge 2} \Big( \frac{\partial \eta }{\partial  x_{i}}  \Big)^{2} \bigg] + ...
        \end{aligned}
    \end{equation}
where $const=\text{Area}(\Sigma\left(A_1)\right)$ and ... denotes the higher order terms of $\eta$. It is equivalent in form as that of a free massless scalar field: 
    \begin{equation}
        S[\eta] = \int_{\Sigma(A_1)} d^{d-1}{x} \sqrt{g} \left( \frac{1}{2}g^{\mu \nu} \partial_{\mu}\eta(x) \partial_{\nu}\eta(x)  \right) = \int_{\Sigma(A_1)} d^{d-1}{x}\sqrt{g}\;\left( \frac{1}{2}\eta(x)\; \Box_{g} \eta(x)\; \right) \nonumber
    \end{equation}
defined on a Euclidean background with the metric $g$:
    \begin{equation}
        ds^{2} = \frac{1}{z^{\frac{2d-2}{d-3}}} \bigg( dz^{2} + dx_{2}^{2} + \cdots + dx_{d-1}^{2} \bigg)
    \end{equation}
and so has the following kinetic operator: 
\be 
\sqrt{g}\;\Box_g =  (d-1) z^{-d} \partial_z-z^{1-d} \partial_z^2-z^{1-d} \partial^2_i
\ee
Notice that the effective metric $g$ for $S(\eta)$ differs from the induced metric $h$ on $\Sigma(A_1)$, which is simply $\text{AdS}_{d-1}$. The kernel is just the Green's function of massless scalar field on this $ d-1 $ dimension curved space which solves the following differential equation:
\be\label{eq:kernel_diff}
\square_{g}  \mathcal{K}\left(z, x_i;z^{\prime},x^{\prime}_i\right)= g^{-1/2}\delta\left(z-z^{\prime}\right)\delta^{d-2}(x_i-x^{\prime}_i)
\ee
In addition, we need to impose the boundary condition that $\mathcal{K}$ vanishes towards the asymptotic boundary $\partial A$: 
\be
\lim_{\vec{x}\to \partial A}\mathcal{K}(\vec{x};\vec{x}')\to 0  
\ee
For $d = 2$ this is an ordinary differential equation, it is therefore straightforward to obtain the Green's function via the Wronskian method. For $ d > 2$, we can exploit the translation symmetry along $x_{i\geq 2}$ to solve (\ref{eq:kernel_diff}) via the Wronskian in the Fourier basis. We leave the details in appendix \ref{The kernel} and summarize the general results below:
\bea\label{eq:kernels}
&& \mathcal{K}(z, z^{\prime}) = \frac{1}{2} \bigg( \Theta(z^{\prime} - z) z^{2} +  \Theta(z - z^{\prime} ) (z^{\prime})^{2}\bigg), \quad d=2\nonumber\\
&& \mathcal{K}\left(z, x_{i} ; z^{\prime}, x_{i}^{\prime} \right) = C_{d} \; z z^{\prime} \; (\xi)^{d-1} \; _2F_1\left(\frac{d -1}{2}, \frac{d}{2} ; \frac{d + 3}{2} ; \xi^{2}\right) , d> 2 \nonumber\\
&& C_{d} = 2^{-(d+1/2)} \pi^{1-d/2}\frac{ \Gamma(d-1)}{\Gamma(d/2+3/2)} ,\quad \xi = \frac{2 z z^{\prime}}{ z^{2} + (z^{\prime})^{2} + \sum_{i = 2}^{d-1} (x_{i} - x_{i}^{\prime})^{2}} 
\eea 

\vspace{2mm}

\subsection{Perturbative causal shadows: explicit results in $\text{AdS}_3$ and $\text{AdS}_4$}
Now we have all the ingredients to compute the shape of the causal shadow $\zeta$ explicitly via (\ref{eq:QES_1st}) by plugging in the results (\ref{var-MI-final}) and (\ref{eq:kernels}). For easy reference we collect them together below in Rindler coordinates:
\begin{equation}
       \zeta\left(z_{1}, x_{i}\right)=4 G_{N}  \int_{0}^{+\infty} {d z_{2}}  \int {d y_{j \ge 2}}  \;\;  \mathcal{K}\left(z_{1}, x_{i} ; z_{2}, y_{j} \right) \left[\frac{\delta I_{a_1\cup a_2}}{\delta \eta\left(z_{2}, y_{j}\right)}\right]+\mathcal{O}\left(G_{N}^{2}\right)
   \end{equation}
where the linear response of the bulk mutual information is: 
\bea
   \frac{\delta I(a_{1}, a_{2})}{\delta \eta(z,x^i)} &=&    N_{d, \Delta}  z^{2\Delta-d}   (1+z^{2}+\sum_{i \geq 2} x_{i}^{2})^{-2\Delta-1} \nonumber\\
       N_{d, \Delta} &=&    \lambda^{2 \Delta }  \frac{ \pi^{3/2} }{4^{2\Delta} \pi^{d/2} } \frac{\Gamma( 2\Delta  +1 )^{2}}{\Gamma(\Delta )  \Gamma(2\Delta + 3/2)} \frac{1}{\Gamma(\Delta -d/2 + 1)}
\eea
and the integration kernel is:
\bea
&& \mathcal{K}(z, z^{\prime}) = \frac{1}{2} \bigg( \Theta(z^{\prime} - z) z^{2} +  \Theta(z - z^{\prime} ) (z^{\prime})^{2}\bigg), \quad d=2\nonumber\\
&& \mathcal{K}\left(z, x_{i} ; z^{\prime}, x_{i}^{\prime} \right) = C_{d} \; z z^{\prime} \; (\xi)^{d-1} \; _2F_1\left(\frac{d -1}{2}, \frac{d}{2} ; \frac{d + 3}{2} ; \xi^{2}\right) , d> 2 \nonumber\\
&& C_{d} = 2^{-(d+1/2)} \pi^{1-d/2}\frac{ \Gamma(d-1)}{\Gamma(d/2+3/2)} ,\quad \xi = \frac{2 z z^{\prime}}{ z^{2} + (z^{\prime})^{2} + \sum_{i = 2}^{d-1} (x_{i} - x_{i}^{\prime})^{2}}  
\eea
Before getting more explicit, one can already infer that $\zeta(x)\geq 0$ is positive-definite. This follows from: (i) the positivity of the linear response of the mutual information, i.e. $\delta I_{a_1\cup a_2}/\delta \eta \geq 0$ for $\eta>0$, which has to hold due to the general monotonicity of the mutual information under shape deformations; (ii) the positivity of all the matrix-elements (not just the spectrum) of the kernel (\ref{eq:kernels}), which can be checked explicitly. Therefore the quantum correction does move the location of the Q.E.S more towards the interior of the bulk than the RT surface, and thus generating a well-defined causal shadow as one expects. 

To recover the results in the spherical coordinates $\tilde{\zeta}(Y)$, one needs to invert the isometric transformation (\ref{eq:map}):
\bea \label{eq:map}
&&  \Lambda x^{\alpha}(Y) = \frac{(Y^{\alpha} - 2R_{1}^{2}C^{\alpha}) + C^{\alpha} (Y - 2R_{1}^{2} C )^{2}}{1 + 2C \cdot (Y - 2R_{1}^{2} C ) + C^{2} (Y - 2R_{1}^{2} C )^{2} } \nonumber\\
\eea
and take into account the transformation property of $\zeta$ as the contra-variant component of a vector field normal to $\Sigma(A_1)$:  
\be
\tilde{\zeta}(Y) = \zeta\left(x(Y)\right)\left(\frac{\partial Y^r}{\partial x^1}\right)\bigg |_{x(Y)}
\ee
where $Y^r=\sqrt{Y_0^2+...Y_d^2}$ is the radial coordinate in $Y$ coordinates whose one-form:
\be
dY^r= d|Y| = R_1^{-1}\left(Y^0 dY^0 + ... +Y^d dY^d\right)
\ee
is the normal to the hemi-sphere $|Y|=R_1$. 

For illustration, we compute two explicit examples in $\text{AdS}_3$ and $\text{AdS}_4$ respectively. The causal shadow for AdS$_{3}$ involves only a single integral over $z$, in which the kernel is a simple step function in (\ref{eq:kernels}). As a result, a closed-form expression can be obtained for $\zeta(x)$ in $\text{AdS}_3$ via the hypergeometric functions: 
\bea\label{eq:CS_AdS3}
\zeta\left(z \right) &=& \frac{ G_N   N_{d, \Delta}}{\Delta+1}\; \bigg[ z^{2\Delta + 2}\; _2F_1\left(\Delta +1, 2\Delta +1 ; \Delta + 2; -z^{2}\right) \nonumber\\
&+&    z^{-2\Delta}\;  _2F_1\left(\Delta +1, 2\Delta +1 ; \Delta + 2; -\frac{1}{z^{2}}\right)  \bigg] +  \mathcal{O}\left(G_{N}^{2}\right)
\eea
We note that $\zeta(z)\sim z^2\to 0$ as $z\to 0$. In the case of $\text{AdS}_3$, the mapping between the Rindler coordinate $z$ and the angular variable $\theta = \sin^{-1}{(Z/R_1)}$ along the semi-circular $\Sigma(A_1)$ in $Y$ coordinates, and the corresponding transformation between the displacement vector fields are given by:
\be\label{eq:map_AdS3}
 z = \frac{ 2 R_{1} \sin \theta }{\Lambda(1 + \cos\theta) }  , \;\theta \in (0, \pi),\;\;\tilde{\zeta}(\theta) = \zeta\left(z\right) \frac{ 4 \Lambda R_{1}^{2} }{  4 R_{1}^{2} + \Lambda^{2} z^{2}}
\ee

\begin{figure}[h]
      \centering
      \includegraphics[scale=0.55]{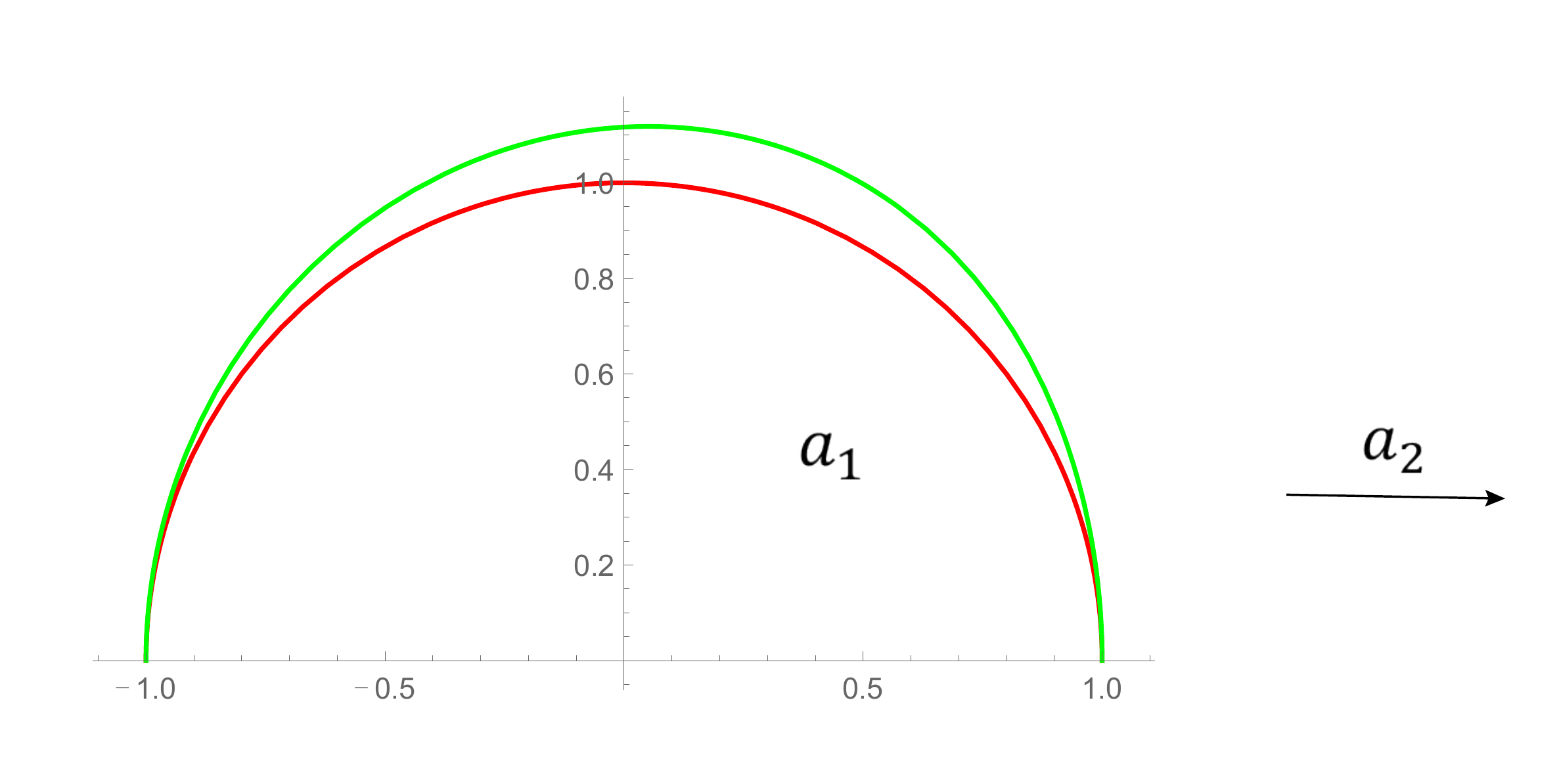}
      \caption{The profile of the causal shadow in terms of the causal horizon (red) against the Q.E.S (green) in AdS$_{3}$, with $ \Delta = 2 $. For illustration purpose we exaggerated the value of $G_N \lambda^{2\Delta}$ to be order 1.}
      \label{QES-AdS3-case}
  \end{figure}

The resulting profile for the causal shadow $\tilde{\zeta}$ can then be obtained from (\ref{eq:CS_AdS3}) and (\ref{eq:map_AdS3}). We can then plot $\tilde{\zeta}$ on top of the un-deformed semi-circular causal horizon $\Sigma(A_1)=\mathcal{C}(A_1)$ to get the Q.E.S, this is to the leading order in both $G_N$ and $\lambda$, see Fig \ref{QES-AdS3-case} for an explicit example. Only the component homologous to $A_1$ is shown, the other component is located as the arrow points.
 
\begin{figure}[h]
      \centering
      \includegraphics[scale=0.50]{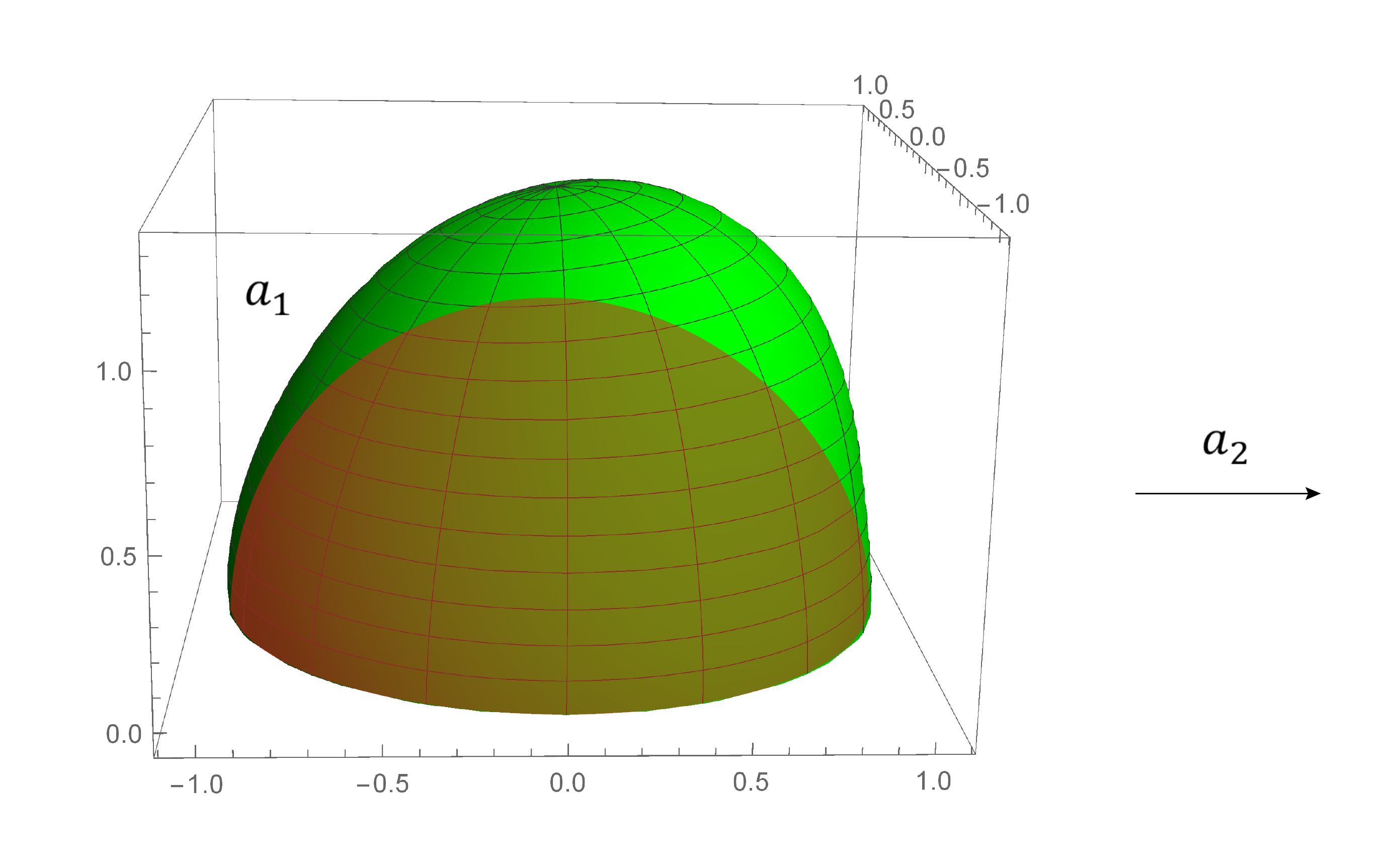}
      \caption{The profile of the causal shadow in terms of the causal horizon (red) against the Q.E.S (green) in AdS$_{4}$, with $ \Delta = 2 $. Again for illustration purpose we exaggerated the value of $G_N \lambda^{2\Delta}$ to be order 1 }
      \label{QES-AdS4-case}
  \end{figure}

Next we consider the two spheres in AdS$_{4}$, the deformation vector field in Rindler coordinates takes the explicit form of a double-integral over as below:
\bea\label{eq:CS_AdS4}
\zeta\left(z, x_2\right)&=&16 G_{N} C_d N_{d, \Delta} \int_0^{+\infty} d z' \int d y_2 \; \left(z z'\right) \left(\frac{\sqrt{1-\xi^2}-1}{\xi}\right)^2\nonumber\\
&\times&\left(z'\right)^{2 \Delta-2}\left(1+\left(z'\right)^2+y_2^2\right)^{-2 \Delta-1}+\mathcal{O}\left(G_{N}^2\right)\nonumber\\
\xi &=& \frac{2z z'}{z^2+(z')^2+(y_2-x_2)^2}
\eea
The map between $(z,x_2)$ and the angular variables $(\alpha,\theta)$ on the hemi-sphere defined by: 
\be
Y_{1}  = R_{1} \cos \alpha \cos \theta, \quad Y_{2} = R_{1} \sin \alpha \cos \theta,\quad  Z = R_{1} \sin \theta, \quad \alpha \in (0,2\pi), \quad \theta \in (0, \pi/2) \nonumber
\ee
together with the transformation between $\zeta$ and $\tilde{\zeta}$ are given explicitly by:
  \bea
    &&  x_{2} = \frac{ 2 R_{1} \sin \alpha \cos \theta }{\Lambda (1+  \cos\alpha \cos \theta )}, \quad   z = \frac{ 2 R_{1} \sin \theta }{\Lambda (1+ \cos\alpha \cos \theta )}\nonumber\\
   && \tilde{\zeta}(\theta,\alpha) = \zeta\left(z, x_{2}\right) \frac{ 4 \Lambda R_{1}^{2}}{  4 R_{1}^{2} + \Lambda^{2} (x_{2}^{2} + z^{2}) }
  \eea
Unfortunately in this case the double integral in (\ref{eq:CS_AdS4}) cannot be integrated to a closed-form expression like the $\text{AdS}_3$ case in any obvious way. Therefore we rely on numerically computing the profile in particular realizations of the parameters, and plot the results of $\tilde{\zeta}$ on top of the un-deformed hemi-spherical causal horizon $\mathcal{C}(A_1) = \Sigma(A_1)$, see Fig \ref{QES-AdS4-case}.

We end the section by the following observation. From the results of both plots for the $\text{AdS}_3$ (Fig \ref{QES-AdS3-case}) and $\text{AdS}_4$ (Fig \ref{QES-AdS4-case}), it is curious to see that the leading order quantum corrections in $G_N$ and $\lambda$ create a tendency for the Q.E.S to tilt towards the other component. In fact for the case $\text{AdS}_3$ one can explicitly compute the angular coordinate $\theta_{peak}$ at which the peak of $\tilde{\zeta}(\theta)$ occurs: 
\bea
\theta_{peak}&=&\frac{\pi}{2}-   4^{\Delta}(\Delta+1/2)K(\Delta)\;\lambda + \mathcal{O}(\lambda^2)\leq \frac{\pi}{2}\nonumber\\
K(\Delta)&=&\frac{{ }_2 F_1(\Delta+1,2 \Delta+1 ; \Delta+2 ;-1)}{\Delta+1}- \frac{{ }_2 F_1(\Delta+2,2 \Delta+2 ; \Delta+3 ;-1)}{\Delta/2+1}
\eea
which indeed is pointing towards $a_2$. Furthermore, the extent of the tilting grows together with the size of the causal shadow as $\lambda$ increases. Such behaviors are as if preparing for the entanglement phase transitions as we tune $\lambda$ across $\lambda_{c}= 1$, after which the entanglement wedges become connected. In the future it will be interesting to study if one can learn more about the non-perturbative phase transition by pushing the perturbative analysis for the disconnected phase $\lambda\ll 1$ to higher orders in $G_N$, via some forms of resurgence analysis.

\vspace{5mm}
  
\section{Aspects of perturbative causal shadow: positivity and more}\label{sec:comments}
In section (\ref{sec:main}) we have done an explicit computation of the geometric shape of causal shadow, it arises from quantum corrections to the Q.E.S relative to the classical RT surface due to bulk mutual information between two distant spheres. As pointed out before, this corresponds to a particular case of perturbative genesis for the causal shadow -- a convenient case in terms of doing computations. In this section we unpack some aspects of the causal shadow perturbation theory from slightly more general perspectives. 

The size of the causal shadow $\delta{\Sigma} = \zeta$ is kept small at leading order by factors of both $G_N$ and $\lambda^{2\Delta}$. We should first comment on the nature of such a perturbatively small region. In the face of quantum gravity in the bulk, the explicit form of the causal shadow shape could be taken with a grain of salt because it may presumably be ``blurred" by the quantum fluctuations of the metric.\footnote{The fluctuations due to the bulk gravitons are of order $G_N^{1/2}$, which then dominate over $\zeta$. We neglect the graviton fluctuations as was also done in the discussion of \cite{Engelhardt2015}.} At the leading order, the causal shadow is built up additively from each OPE contributions of the bulk mutual information: 
\be
\zeta =G_N \sum_{\Delta'} \lambda^{2\Delta} \tilde{\zeta}_{\Delta'}+\mathcal{O}(G_N^2), \;\; I_{\text{bulk}}= \sum_{\Delta'} \lambda^{2\Delta'} I_{\Delta'} 
\ee
Therefore in the case of large number $N$ of light fields in the effective bulk QFTs, the perturbative causal shadow $\zeta \propto N G_N$ could still contrast well against metric fluctuation. In our case, the classical RT surfaces are special sub-manifolds protected by bulk isometries -- they are the invariant surfaces. As a result, the deformation away from them, even of quantum nature, is more revealing than in generic cases. It is worth reminding that at order $G^1_N$, the bulk entanglement entropy (and thus the generalized entropy) itself also receives contributions from the interactions of the bulk fields, though such corrections do not affect the location of the Q.E.S at $G^1_N$. Related to this point, one might wonder if the correction $\zeta \propto G_N \lambda^{2\Delta}$ to the Q.E.S affects the evaluation of the generalized entropy $S_{gen}(a_1\cup a_2)$ in a way that agrees with the boundary CFT result. Setting $\zeta=0$, the bulk and boundary mutual information has been shown in \cite{Faulkner2016Aug} to agree at the leading order in OPE: $I_{bulk}=I_{CFT}\sim \lambda^{2\Delta}$. The effect from $\zeta$ on $S_{gen}(a_1\cup a_2)$ is expected to be of order $G_N \lambda^{4\Delta}$. From the bulk perspective, one needs to consider other $G^1_N$ effects, e.g. from the bulk interactions; from the CFT perspective, probing such effects would require the computation to be pushed to the next order in both $1/N$ and OPE. It will be interesting to pursue these in future investigations \footnote{We thank the referee for bringing up the last two points.}.

\subsection{Perturbative causal shadow and positivity conditions in boundary CFTs}
The most important aspect regarding the deformation in our case is its positivity, i.e. $\zeta(Y)\geq 0$ for all $Y\in \Sigma_{RT}$. This is a direct manifestation that the deformation in this case represents a causal shadow, and so must be positive definite. Technically the positivity follows from the combination of the monotonicity for bulk mutual information under shape deformation, and the positive definiteness of the kernel function related to the second-variation of the area-functional. More generally, the notion of causal shadow makes sense only because there is a ``positive-definite" relation between the position of Q.E.S and that of the causal horizon -- the former always lies more to the interior than the latter. In general bulk terms, regarding the classical RT surfaces this is a consequence of the null energy condition (NEC), which is related to the strong subadditivity (SSA) constraint in the boundary CFTs; at a more refined level regarding the Q.E.S, it is a consequence of the bulk focusing conjecture.  

In our case, due to the coincidence between the Q.E.S and the causal horizon before perturbation, the positive-definiteness transcends into perturbative theory and is sharpened into a class of local positivity conditions defined on the un-deformed Q.E.S. In principle, they should reflect properties of the deformation as probed by the local modular flow before perturbation. Their meanings will be more interesting from the perspective of boundary CFTs. In \cite{Faulkner:2017} it has been found that for a holographic state $\psi$ and boundary subregion $A$ whose modular Hamiltonian we simply write as $H_0$, by computing correlators involving the so-called modular zero-mode, bulk locality near the corresponding RT surface $\Sigma_{RT}$ can be probed with the help of bulk-boundary correlators $\langle \Phi \mathcal{O} \rangle$:
\be \label{eq:FL_corr}
\int^\infty_{-\infty} ds \left\langle e^{iH_0 s} \mathcal{O}(x) e^{-iH_0 s}\mathcal{O}(y)\right\rangle_\psi = 4\pi \int_{\Sigma_{RT}}\sqrt{h}\; dY_{RT} \langle \Phi(Y_{RT})\mathcal{O}(x) \rangle\;\langle \Phi(Y_{RT})\mathcal{O}(y)\rangle
\ee    
for $x,y \in D(A)$. Now one can imagine deforming the modular Hamiltonian by a small correction: 
\be 
H_0\to H'=H_0+\Delta H
\ee
and expand the LHS of (\ref{eq:FL_corr}) to extract the linear order in $\Delta H$ term:
\bea\label{eq:FL_1st_L}
&i& \int^\infty_{-\infty}ds \int^{s}_0 dt\;\Big\langle e^{iH_0s}\left[e^{-iH_0t} \Delta H e^{iH_0t},\mathcal{O}(x)\right] e^{-iH_0s} \mathcal{O}(y) \Big\rangle_\psi \nonumber\\
&=& i\int^\infty_{-\infty}ds \int^{s}_0 dt\;\Big\langle \left[\Delta H ,\mathcal{O}_t(x)\right]  \mathcal{O}_{t-s}(y) \Big\rangle_\psi
\eea
where $\mathcal{O}_t(x)=e^{iH_0 t}\mathcal{O}(x) e^{-iH_0t}$ is the modular flowed local operator using $H_0$. Notice that in doing this we are still evaluating the correlator in the original state $\psi$. A deformation $\Delta H$ of generic nature will place (\ref{eq:FL_1st_L}) beyond the context of (\ref{eq:FL_corr}). This is because $\Delta H$ may involve an actual change of the state $|\psi\rangle \to |\psi'\rangle$, while (\ref{eq:FL_corr}) holds only when the LHS is evaluated in the defining state of $\tilde{H}$. We can choose to focus on $\Delta H$ that arises not as a result of changing the state, but instead from changing the subregion $A\to A'$, for the moment we assume to be purely spatial. In this case (\ref{eq:FL_1st_L}) captures the leading order change of the RT surface on the RHS:
\be
\Sigma_{RT}\to \Sigma'_{RT}
\ee
In this case both $\Sigma_{RT}$ and $\Sigma'_{RT}$ are in the same bulk geometry. If the subregions $A$ and $A'$ across deformation are homologous, i.e. there is a continuous one-parameter family of subregions interpolating between $A$ and $A'$, then one can assume the same about the RT surfaces (we are of course excluding entanglement phase transitions in between), and write $\Sigma'_{RT}= \Sigma_{RT}+\zeta$. In this way one can expand the RHS of (\ref{eq:FL_corr}) to extract the linear order term in $\zeta$, and matching it with the linear order (\ref{eq:FL_1st_L}) obtain:  
\bea\label{eq:FL_1st}
 i\int^\infty_{-\infty}&ds& \int^{s}_0 dt\;\Big\langle \left[\Delta H ,\mathcal{O}_t(x)\right]  \mathcal{O}_{t-s}(y) \Big\rangle_\psi = 4\pi \int_{\Sigma_{RT}}\sqrt{h}\; dY_{RT} K_N(x,y;Y_{RT}) \zeta(Y_{RT})\nonumber\\
&&K_N(x,z;Y_{RT})= \partial_{N}\Big(\left\langle \Phi(Y_{RT})\mathcal{O}(x) \right\rangle \left\langle \Phi(Y_{RT}) \mathcal{O}(z) \right\rangle\Big)  
\eea
where $\partial_{N}$ represents the bulk spatial derivative normal to $\Sigma_{RT}$ at $Y_{RT}$, pointing towards the interior. In some sense (\ref{eq:FL_1st}) provides a dictionary up to linear order that identifies between the operator $\Delta H$ in the boundary CFT with the geometric data $\zeta$ in the bulk. In more precise terms, the LHS is a convoluted integral of the $\Delta H$ matrix elements between states created by the un-deformed modular-flowed operators. With explicit knowledge of $\Delta H$ and in the specific case of our interest where the modular flow is local it is plausible at least in principle to compute it explicitly. As an exercise, in Appendix (\ref{app:check}) we explicitly compute the LHS for the deformed half-space in the vacuum state and verify (\ref{eq:FL_1st}) in this case. The RHS is a bulk integral over the RT surface of the normal displacement field $\zeta(Y)$ dressed to the boundary by the kernel $K_N(x,z;Y_{RT})$, which contains information about the original RT surface and bulk geometry through the bulk-boundary propagators, i.e. $H_0$ and $\psi$. The dictionary (\ref{eq:FL_1st}) is defined for arbitrary $x,y\in D(A)$, one can make better use of them by integrating over $\lbrace x,y\in D(A)\rbrace$ with appropriate smearing functions $f(x,y)$.

If the spatial deformation $A\to A'$ is such that $A'\supseteq A$, it is expected that $\zeta(Y_{RT})\geq 0$ for all $Y_{RT} \in \Sigma_{RT}$. This is known as the entanglement wedge nesting condition and in this case constitutes one of the positivity conditions regarding $\Delta H$ through the dictionary (\ref{eq:FL_1st}). It holds for generic $H_0$ with a holographic dual and reflects the monotonicity of the bulk reconstruction capacities under subregion nesting. The notion of causal shadow gives rise to a different positivity condition regarding $\Delta H$, it requires that under generic change of subregion $A\to A'$ but for the special cases of our interest where $H_0$ drives a local modular flow, the corresponding normal displacement field $\zeta$ satisfies that: 
\be \label{eq:cs_ps}
\zeta(Y_{RT}) \geq \chi(Y_{RT})
\ee
for all $Y_{RT}\in \Sigma_{RT}$, where $\chi$ is the corresponding normal displacement field of the deformed causal horizon towards the interior: 
\be
\mathcal{C}_A \to \mathcal{C}_{A'} = \mathcal{C}_A+\chi 
\ee 
However, in this case it is difficult in general to obtain a characterization of this positivity condition in boundary CFTs. The main reason is that we do not yet have objects in the boundary CFT that captures the location of causal horizon like (\ref{eq:FL_corr}) does, and thus unable to fully reconstruct all ingredients involved in the positivity constraint (\ref{eq:cs_ps}).  

We can nevertheless speculate about the origin of these perturbative positivity constraints related to the genesis of causal shadow. Roughly speaking it is a result of the additional bulk reconstruction capacities that are gained from exploiting the entanglement structure of $\rho_A$, i.e. the modular flow, compared with that of the HKLL methods which uses the emergent causal structure of the bulk state. In particular, the latter only probes the causal structure seen by local operators; while the former probes a more subtle notion of causality due to its non-local property in general. In other contexts, viewing the notion of causality through the lens of the modular flow has also allowed one to discover deeper aspects of QFTs, e.g. proving the quantum null energy conditions (QNEC) \cite{HWang2019Sep, Faulkner1804}. In the special case where the modular flow is local, the entanglement structure reflects nothing more than the local notion of causality itself. As a result the two bulk reconstruction approaches are equivalent. There is no additional mileage gained from using the entanglement structures, and the causal shadow is thus degenerate. By perturbing away from these special cases, one might be able to capture a notion of monotonicity regarding the extent to which the deformed modular flow is local -- at a heuristic level it simply cannot become more local than the un-deformed one. It could still be the case that the deformed modular flow is local but with a different set of modular trajectories, e.g. from a sphere $A$ to another sphere $A'$. The positivity constraint (\ref{eq:cs_ps}) may be viewed as a reflection of this monotonicity. 

\subsection{Entanglement wedge nesting from correlations}
We now focus our attention more on the particular case of two distant spheres. Recall that the computation of the causal shadow profile $\zeta$ in section (\ref{sec:main}) amounts to the quantum correction to the location of the Q.E.S relative to the classical RT surface, which coincides with the causal horizon. As has been set up in (\ref{sec:main}), one should view this as the response of the Q.E.S from the following deformation in modular Hamiltonian from that of two un-correlated spheres: 
\be
H_0 = -\ln{\rho_{A_1}}\otimes \mathds{1}_{A_2} - \mathds{1}_{A_2} \otimes \ln{\rho_{A_1}}
\ee
to that of the correlated spheres: 
\be
H(\lambda) = -\ln{\left(\rho_{A_1\cup A_2}\right)} = H_0 + \Delta H(\lambda) 
\ee
where the conformal ratio $\lambda = 4 R_1 R_2/L^2\geq 0$ parametrizes the deformation away from the un-correlated limit $\lambda = 0$. 

It then allows the possibility of understanding the positivity constraint (\ref{eq:cs_ps}) better because in this case it simply means a positive-definite shift of the location of Q.E.S under the deformation $\Delta H(\lambda)$. Unfortunately such a shift is not captured by (\ref{eq:FL_1st}) at the leading order in large $N$. A naive application of (\ref{eq:FL_1st}), say with both $\mathcal{O}(x)$ and $\mathcal{O}(y)$ inserted in the spherical causal domains $D(A_1)$, will yield on the RHS a response that simply accounts for the effect of the other RT surface $\Sigma_{A_2}$ moving from infinity to finite $L$, or conformal equivalently from being point-like to a hemi-sphere with finite radius of $R_2$: 
\be
i\int^\infty_{-\infty}ds \int^{s}_0 dt\;\Big\langle \left[\Delta H(\lambda) ,\mathcal{O}_t(x)\right]  \mathcal{O}_{t-s}(y) \Big\rangle = 4\pi \int_{\Sigma_{A_2}}\sqrt{h}\; dY \langle \Phi(Y)\mathcal{O}(x)\rangle \langle \Phi(Y)\mathcal{O}(y)\rangle\nonumber\\ 
\ee
through which we gain no access to the positivity constraint $\zeta\geq 0$. This is not surprising because $\zeta\propto G_N$ corresponds to a sub-leading in $1/N$ effect in boundary CFTs. One would need to go beyond the wick-contraction when computing the modular-flowed correlators in the LHS to capture this. 

Let us contemplate on the origin of the positivity $\zeta\geq 0$ in this particular case through properties of $\Delta H(\lambda)$ in the boundary CFTs. On the one hand, it is a special case of the general positivity condition (\ref{eq:cs_ps}) for $\chi=0$ and so is related to the non-local aspects of the modular flow caused by the deformation $\Delta H$. On the other hand, it also takes the form of a positive-definite shift of the Q.E.S before and after the deformation, so is analogous to the statement of the entanglement wedge nesting. From the perspective of the subregion $A_1 \cup A_2$, the entanglement wedge expands not because the subregion is geometrically enlarged at $\lambda >0$, but because $\rho_{A_1\cup A_2}(\lambda)$ includes correlation between the two components $A_1$ and $A_2$ at $\lambda >0$: 
\be
\rho_{A_1\cup A_2} =\rho_{A_1} \otimes \rho_{A_2} + \delta \rho_{A_1,A_2}(\lambda) 
\ee 
At the level of the reduced density matrix, the deformation $\delta\rho_{A_1,A_2}(\lambda)$ cannot be arbitrary in our case, it needs to be constrained such that: 
\be\label{eq:MI_constraint}
\text{Tr}_{A_2}\left(\rho_{A_1\cup A_2}\right) = \rho_{A_1},\;\; \text{Tr}_{A_1}\left(\rho_{A_1\cup A_2}\right) = \rho_{A_2}
\ee
This aspect of $\delta \rho_{A_1,A_2}(\lambda)$ makes it behave in ways analogous to the context of the entanglement wedge nesting. In some sense the correlated $\rho_{A_1\cup A_2}$ contains a positive-definite amount of more information than $\rho_{A_1}\otimes \rho_{A_2}$ due to $\delta \rho$ -- one can obtain those of the latter by taking appropriate partial trace on the former. A natural consequence of (\ref{eq:MI_constraint}) is the positivity of the mutual information, however as an operator statement  its content should be richer. In particular by taking operator logarithms on both sides one can derive constraints regarding the modular Hamiltonian deformation $\Delta H(\lambda)$, it is possible that the origin for the positivity of $\zeta$  in the boundary CFT is encoded in such constraints. 

\subsection{Causal structure v.s. entanglement structure}
In more general terms, the notion of the causal shadow comes from comparing the causal structure against the entanglement structure in AdS/CFT. We finish this section by making some further comments regarding this. 

From the perspective of the boundary CFTs, the entanglement and causal structures are in general controlled by different ingredients. While the entanglement structure concerns the properties of the particular state $\psi$; the causal structure concerns the theory itself, in particular the background spacetime it is defined upon. Admittedly these two aspects may not be completely independent, as the characterization of sensible states reflects the dynamics of the underlying theory. For example, by restricting to the vacuum $\Omega$ or the nearby low-energy states we probe the long wavelength aspects of the theory. In the context of algebraic QFTs, the entanglement structure should be thought of as a property of both the state and the observable operator algebra of the theory \cite{Witten_2018_RMP}.

In terms of the Schwinger-Keldysh path-integral, the state $\psi$ can be prepared by a Euclidean section with appropriate sources inserted; the dynamical evolution of the theory including the causal structure is reflected on the subsequent Lorentzian section that is glued to the state-preparing Euclidean section. In the classical analysis of section (\ref{sec:degenerate}) regarding local modular flows, we assumed that the two sections can be related by a simple wick rotation $s\to i\theta$. The existence and simplicity of such a wick rotation is a result of the conformal isometries responsible for the local property of the modular flow. In generic situations neither should there exist, nor should the Lorentzian and Euclidean sections be related by such a convenient wick rotation. 

In the context of AdS/CFT, the entanglement and causal structures are more closely related in the bulk. This is mainly because the bulk geometry, in particular the associated causal structures, emerge dynamically from the boundary data including both the state $\psi$ and the CFT Hamiltonian. At the leading order in $1/N$, it is obtained by solving the bulk gravitational equations subject to the boundary condition of $\psi$, e.g. imposed on some bulk initial-slice that is space-like; and that of the CFT Hamiltonian, imposed along the time-like asymptotic boundary. The Q.E.S is fixed locally by the spatial slice on which the bulk wave-functional can be defined; while the causal horizon is obtained only after one finds the bulk Lorentzian geometry that evolves to the future of the spatial slice. Their relation in terms of the causal shadow is therefore a more tangled-up reflection of the relation between the entanglement and causality in the boundary CFTs. 

\section{Summary and outlook}\label{sec:discuss}   
In this paper, we discussed the relationship between causal shadows and the properties of modular flows. We analyzed in section (\ref{sec:degenerate}) the cases where the corresponding modular flow is local, and tracked the essential reasons behind the degeneracy of the causal horizon based on the known examples but discussed in more general context. The existence of the boundary conformal isometries extendable into the bulk severely constrains the caustics structure on the orthogonal null congruences, making it necessary for the RT surface and the causal horizon to coincide; it also constrains the quantum corrections so that it does not modify the location of the Q.E.S. 

We then studied in section (\ref{sec:bulk_MI}) and (\ref{sec:main}) the genesis of a non-degenerate causal shadow under a perturbation that destroys the conformal isometries controlling the local property of the modular flow. We focused on the perturbation theory involving two spheres $A_1$ and $A_2$ in the vacuum separated by large distance $L$, treating the correlation at finite $L$ as the perturbation to the $L=\infty$ case. The causal shadow is generated as a quantum correction to the location of the Q.E.S arising from the mutual information $I_{A_1,A_2}$ whose OPE can be obtained term by term in the large $L$ limit. At the leading order in $G_N$, we solved the Q.E.S equation perturbatively near the classical RT surface, which is the same as the causal horizon, and computed a typical OPE contribution from a primary scalar to the shape $\zeta$ of the causal shadow. The profile $\zeta$ on both spheres is found to be positive-definite and bending towards each other. Explicit results are presented for the case of two spheres in $\text{AdS}_3$ and $\text{AdS}_4$.

We provided in section (\ref{sec:comments}) some general comments concerning properties of the perturbative causal shadows from the perspective of boundary CFTs. We discussed the possibility and issues of extracting positivity conditions related to the causal shadow using the boundary modular zero-modes in the context of the bulk reconstruction on the RT surfaces; we also drew an analogy between the causal shadow we computed in section (\ref{sec:main}) and the statement of entanglement wedge nesting, but from correlations instead of from subregion enlarging. 

We conclude the paper by mentioning some future directions. Firstly, it is interesting to study other cases of perturbative causal shadows by allowing various types of perturbations, e.g. deforming the sub-region geometry or deforming the underlying states. As has been discussed, in generic cases not only the Q.E.S but also the causal horizon shifts. While generic Cauchy horizons and event horizons are plagued with caustics and thus are hard to analyze in full details, computing their linear responses away from the smooth cases is plausible, such as in \cite{Levine:2020upy}. It is also important to think about objects in the boundary CFTs that probe the location of the causal horizon like the modular zero-mode does for the RT surface. Finding such objects is a necessary step towards understanding from the boundary CFT perspectives the positivity conditions (\ref{eq:cs_ps}) of causal shadows in more generic cases; it may also provide additional clues regarding the nature of the holographic causal information \cite{Hubeny2012}. Related to this, it is useful to devise quantitative measures for the extent of non-locality related to the modular flow, e.g. via the operator spreading, etc. They may point to an organizing principle for extracting more information from the positivity condition of the causal shadow. One may also study further the ``dynamical" aspects of the causal shadow by actively probing it from the boundary CFTs, e.g. by applying the so-called Connes cocyle flows \cite{Faulkner1804, Bousso2020} and study the response, something along this line has been done in \cite{Levine:2020upy}. Last but not the least, it is interesting to analyze the holographic entanglement transitions, e.g. in the two sphere case, from the perspective of the causal shadow condensation via non-perturbative effects. 

\section*{Acknowledgments}
We thank Thomas Faulkner and Xinan Zhou for useful discussions. We thank Tadashi Takayanagi and Mukund Rangamani for the reading and the comments on the manuscript. This work is supported by National Science Foundation of China (NSFC) grant no. 12175238.

\appendix

 \section{More details on solving the kernel $\mathcal{K}\left(z_{1}, x_{i} ; z_{2}, y_{j} \right) $} \label{The kernel}
 In the main context, we have defined the kernel $\mathcal{K}$ as the inverse matrix  of the second order variation of the area functional $\operatorname{Area}(\Sigma^{\prime})$, that is: 
  $$
 \mathcal{K}\left(z_{1}, x_{i} ; z_{2}, y_{j} \right) =\left(\left.\frac{\delta^{2} \operatorname{Area}(\Sigma^{\prime})}{\delta \eta\left(z_{1}, x_{i}\right) \delta \eta\left(z_{2}, y_{j} \right)}\right|_{\Sigma^{\prime} =\Sigma_{R T}}\right)^{-1}
 $$
For $d=2$, the kernel satisfies the differential equation: 
\be
\left(z^{-2}\partial_z - z^{-1}\partial^2_z\right) \mathcal{K}(z,z')=\delta (z-z')  
\ee
The two linearly independent solutions to the homogeneous equation are 
\be
f_1(z)= const,\;\;f_2(z) = z^2
\ee
For the one-variable problem, given two linearly independent solutions $\lbrace f_1,f_2\rbrace$ one can use the Wronskian determinant method to construct the Green's function in general: 
\bea
\mathcal{K}(z,z') &=& \frac{1}{W(z)}\Big(\Theta(z-z')f_1(z) f_2(z')+\Theta(z'-z)f_1(z')f_2(z)\Big)\nonumber\\
W(z)&=& f_1(z)f_2'(z)-f_2(z)f_1'(z)
\eea 
The choice for the particular pair of solutions $\lbrace f_1,f_2\rbrace$ is determined by the boundary conditions: 
\be 
\lim_{z\to 0}\mathcal{K}(z,z')\to 0,\;\;\lim_{z\to \infty}\mathcal{K}(z,z')<\infty
\ee 
which fixes the kernel in $d=2$ to be:
\be 
\mathcal{K}(z,z')= \frac{1}{2}\Big(\Theta(z'-z)z^2+\Theta(z-z')(z')^2\Big)
\ee
For $d > 2$, the corresponding differential equation takes the form: 
\be
\left((d-1) z^{-d} \partial_z-z^{1-d} \partial_z^2-z^{1-d} \partial^2_i\right) \mathcal{K}\left(z, x_i ; z^{\prime},y_i\right)=\delta\left(z-z^{\prime}\right) \delta^{d-2}(x_i-y_i)
\ee
The equation has translation symmetry in the $x_i$-plane, so we can go to the Fourier mode basis of $\mathcal{K}\left(z_{1}, x_{i} ; z_{2}, y_{j} \right)=\int d\vec{k}\;e^{i\vec{k}\cdot(x-y)} \mathcal{K}(z_1,\vec{k}; z_2)$, which satisfies: 
 \begin{equation}
     \left((d-1) z^{-d} \partial_z-z^{1-d} \partial_z^2+z^{1-d} k^2\right) \mathcal{K}\left(z, \vec{k} ; z^{\prime}\right)=\delta\left(z-z^{\prime}\right)
 \end{equation}
The remaining steps proceeds the same as the $d=2$ case using the Wronskian determinant method, the solution takes the form as:
 \begin{equation}
     \mathcal{K} \left(z, \vec{k} ; z^{\prime}\right)=\left(z z^{\prime}\right)^{d / 2} \bigg(\Theta\left(z^{\prime}-z\right) I_\nu(k z) K_\nu\left(k z^{\prime}\right)+\Theta\left(z-z^{\prime}\right) K_\nu(k z) I_\nu\left(k z^{\prime}\right)\bigg)
 \end{equation}
 where $ \nu = \frac{d}{2} $ and the boundary conditions are  Dirichlet boundary conditions:
 \begin{equation}
     \left. \mathcal{K} \left(z, \vec{k} ; z^{\prime}\right)\right|_{z=0}=0, \left. \quad \mathcal{K}\left(z, \vec{k} ; z^{\prime}\right)\right|_{z=\infty}=0
 \end{equation}
Then we can Fourier transform the kernel back to the coordinates space as:
\begin{equation}
    \begin{aligned}
        \mathcal{K}\left(z, x_i ; z^{\prime}, x_i^{\prime}\right) =&\; \frac{1}{(2 \pi)^{d-2}} \int d^{d-2} k \; \exp \left(i \mathbf{k} \cdot\left(\mathbf{x}-\mathbf{x}^{\prime}\right)\right) \mathcal{K} \left(z, k ; z^{\prime}\right) \\
        =&\;  C_{d} \; z z^{\prime} \; (\xi)^{d-1} \; F\left(\frac{d -1}{2}, \frac{d}{2} ; \frac{d + 3}{2} ; \xi^{2}\right) 
    \end{aligned}
\end{equation}
where the prefactor $C_{d}$ and the parameter $ \xi $ are given by:
$$ C_{d} = \frac{1}{2^{d+1/2} \pi^{d/2-1}}\frac{ \Gamma(d-1)}{\Gamma(d/2+3/2)} ,\quad \xi = \frac{2 z z^{\prime}}{ z^{2} + (z^{\prime})^{2} + \sum_{i = 2}^{d-1} (x_{i} - x_{i}^{\prime})^{2}} $$ 

\section{Eqn (\ref{eq:FL_1st}) for the deformed half-space in the vacuum}\label{app:check}
In this appendix we try (\ref{eq:FL_1st}) out with some details and verify it in the specific case of the vacuum half-space $A=\lbrace{\t=0,\;x<0,\;y\in \mathds{R}^{d-2}}\rbrace$ in the boundary $\text{CFT}_{d}$. In this case the modular Hamiltonian is simply the Rindler boost generator in half-space: 
\be
H_0 = \int dy\int^{\infty}_0 dx\; x\;T_{tt}(x,y) 
\ee 
We now deform $A$ by introducing a spatial deformation profile $\xi$: 
\be 
A\to A' = \lbrace \t=0, \;x>\xi(y),\; y\in \mathds{R}^{d-2}\rbrace
\ee 
The deformation in modular Hamiltonian to linear order in $\xi$ can be obtained as \cite{HWang2016Sep}: 
\bea\label{eq:MH_HP}
\Delta H = \int dy\;\ \xi^u(y)\int^\infty_{-\infty} du T_{uu}(u,v=0,y) + \int dy\; \xi^v(y) \int^\infty_{-\infty} dv T_{vv}(u=0,v,y)
\eea
where $u=x+t,\;v=x-t$. Spatial deformations correspond to $\xi^u = \xi^v$. If we are instead interested in the case of deformation along the null plane, e.g. $\xi^v=0$, then (\ref{eq:MH_HP}) becomes exact to all orders in $\xi$, thanks to the half-sided modular inclusion property \cite{Casini:2017roe}. At the linear order in $\xi$, we can just focus on checking (\ref{eq:FL_1st}) for one of the null components $\xi^u$ only. In \cite{HWang2016Sep}, it has been explicitly checked in this case that the deformed full modular Hamiltonian agrees between boundary and the bulk and thus realizing what JLSM conjectured: 
\be\label{eq:JLMS}
\Delta H^{\text{full}} = \int_{\Sigma_{RT}}\sqrt{h}\;dz dy\; \zeta^u(z,y) \int^\infty_{-\infty} du\; T^{\text{bulk}}_{uu}(u,v=0,z,y),\;\;Y_{RT}=(z,y)  
\ee 
where $\zeta(z,y)$ is the displacement profile for the deformed RT surface satisfying $\lim_{z\to 0}\zeta(z,y) = \xi(y)$. The agreement was verified explicitly at the level of the bulk free QFT which means leading order in large $N$. In this limit one can write $T^{\text{bulk}}_{uu}$ in terms of bilinears of free fields in the bulk. Therefore to compute its correlation function with single-trace scalar operator $\mathcal{O}$ in the boundary CFT, the stress tensor is effectively just: $T^{\text{bulk}}_{uu} = \partial_u \Phi \partial_u \Phi$. We need to compute its commutator with $\mathcal{O}(x),x\in D(A)$, the half-sided modular Hamiltonian acts the same as the full modular Hamiltonian, so we can use (\ref{eq:JLMS}) but restrict to $u\in[0,\infty]$ whose integral can be converted into a modular integral: 
\be
\int^\infty_{0}du T^{\text{bulk}}_{uu}(u,v=0,Y_{RT})= \int^\infty_{-\infty} ds' e^{-s'} e^{iH_0 s'} T_{uu}(u_0, Y_{RT}) e^{-iH_0s'}\nonumber
\ee
where $u_0>0$ is some arbitrary null regulator away from the RT surface. Now plugging these into the LHS of (\ref{eq:FL_1st}), by re-writing $s'-t \to s'$ and wick contracting among the operators, we obtain the RHS of (\ref{eq:FL_1st}) at leading order in large $N$ as: 
\bea\label{eq:RHS_1}
\text{RHS}&=& 2i u_0\int_{\Sigma_{RT}}\sqrt{h}\;dY_{RT}\; \zeta^u(Y_{RT})\int^\infty_{-\infty}ds (1-e^{-s}) \int^{\infty}_{-\infty} ds' e^{-s'}\Big\langle \left[\partial_u\Phi(u_0,Y_{RT}),\mathcal{O}_{-s'}(x)\right]\Big\rangle\nonumber\\
&\times & \Big\langle \partial_u\Phi(u_0,Y_{RT})  \mathcal{O}_{-s'-s}(y) \Big\rangle
\eea
We can further re-write $s'+s\to s$ but keeping track of the imaginary parts relevant for the operator ordering, after some massaging one obtains two contributions:
\be \label{eq:RHS_2}
\text{RHS} = \int_{\Sigma_{RT}} \sqrt{h} \;dY_{RT}\;\zeta^u(Y_{RT})\;\Big[\mathcal{F}(Y_{RT};x,y) +\mathcal{F}(Y_{RT};y,x)\Big]
\ee
corresponding to the two terms in the $(1-e^{-s})$ factor of (\ref{eq:RHS_1}). The integrand $\mathcal{F}(Y_{RT};x,y)$ takes the following form:
\bea
\mathcal{F} = 2i u_0 \int^{\infty}_{-\infty} ds e^{s}\Big\langle \partial_u\Phi(u_0 e^{s},Y_{RT})  \mathcal{O}(y) \Big\rangle \int^\infty_{-\infty}ds' \Big\langle \left[\partial_u\Phi(u_0 e^{s'},Y_{RT}),\mathcal{O}(x)\right]\Big\rangle
\eea
Using the decaying property of the correlators for large positive $s\to \infty$, one obtains that: 
\bea
&&u_0 \int^{\infty}_{-\infty} ds e^{s}\Big\langle \partial_u\Phi(u_0 e^{s},Y_{RT})  \mathcal{O}(y) \Big\rangle = \int^\infty_0 du \partial_u \langle \Phi(u,Y_{RT})  \mathcal{O}(y) \Big\rangle =- \langle \Phi(Y_{RT})\mathcal{O}(y)\rangle\nonumber\\
&&\int^\infty_{-\infty}ds' \Big\langle \left[\partial_u\Phi(u_0 e^{s'},Y_{RT}),\mathcal{O}(x)\right]\Big\rangle = 2\pi i \langle\partial_{u}\Phi(Y_{RT})\mathcal{O}(x)\rangle
\eea
where we have closed the contour vertically at $\text{Re}\;s'=-\infty$ when evaluating the commutator. Now one can assemble everything in (\ref{eq:RHS_2}) together and verify (\ref{eq:FL_1st}) explicitly for the case of deformed half-space in holographic vacuum: 
\be
\text{RHS} = 4\pi \int  _{\Sigma_{RT}} \sqrt{h} \;dY_{RT}\;\zeta^u(Y_{RT})\partial_u \Big(\langle \Phi(Y_{RT})\mathcal{O}(x) \rangle\langle \Phi(Y_{RT})\mathcal{O}(y) \rangle\Big)
\ee
This by itself also provides a non-trivial perturbative check for the validity of the zero-mode bulk reconstruction formula (\ref{eq:FL_corr}).

\bibliographystyle{JHEP} %using the package of JHEP.
\bibliography{./references/ref}

\end{document}